\documentclass[12pt]{article}
\usepackage{mathtools,amssymb,amsfonts,graphicx}

\usepackage{hyperref}
\usepackage[nosort]{cite}
\usepackage{verbatim}
\usepackage{multicol,color,xcolor,longtable}

\definecolor{darkred}{rgb}{0.65,0.15,0}
\definecolor{darkgreen}{rgb}{.05,.5,.05}
\hypersetup{pdfborder={0 0 0},colorlinks=true,urlcolor=darkred,citecolor=blue,linkcolor=darkred,linktocpage=true}

\usepackage{soul}
\setulcolor{green}

\newcommand{\nn}{\nonumber}

\newcommand{\lb}{\left\{}
\newcommand{\rb}{\right\}}

\makeatletter

\@addtoreset{equation}{section}
\makeatother

\newcommand{\be}{\begin{equation}}
\newcommand{\ee}{\end{equation}}

\begin{document}

\mbox{}

\vspace{20mm}

\begin{center}
  {\LARGE \sc  Integrable deformations of \\[3mm]
  dimensionally reduced gravity}
    \\[13mm]

{\large
Mattia Ces\`aro${}^{1}$ and  David Osten${}^2$}

\vspace{8mm}
${}^1${\it Max-Planck-Institut f\"{u}r Gravitationsphysik (Albert-Einstein-Institut)\\
Am M\"{u}hlenberg 1, DE-14476 Potsdam, Germany}
\vskip 1.2 ex
${}^2${\it Institute for Theoretical Physics (IFT), University of Wroc\l aw,\\
pl. Maxa Borna 9, 50-204 Wroc\l aw, Poland}

\end{center}


\begin{center} 
\hrule

\vspace{6mm}

\begin{tabular}{p{14cm}}
{\small%
Dimensional reduction of gravity theories to $D=2$ along commuting Killing isometries is well-known to be classically integrable. The resulting system typically features a coset $\sigma$-model coupled to a dilaton and a scale factor of the dimensional reduction. In this article, we construct two families of deformations of dimensionally reduced gravity that preserve the Lax integrable structure. 

The first family is an extension of the Auxiliary Field Deformation recently introduced by Ferko and Smith, while the second family consists in the embedding of the Yang-Baxter $\sigma$-model into $D=2$ dimensionally reduced gravity. For both deformations we construct flat Lax representations. The Auxiliary Field Deformation, in particular, preserves the rich algebraic structure underlying the undeformed model and, leaving the  canonical structure of the Lax connection's spatial components essentially unchanged, allows us to prove its integrability also in the Hamiltonian sense.}
\end{tabular}
\vspace{5mm}
\hrule
\end{center}

\hypersetup{pageanchor=false}
\thispagestyle{empty}

\newpage
\hypersetup{pageanchor=true}
\setcounter{page}{1}


\tableofcontents
\newpage

\section{Introduction and summary}
In General Relativity in $D=4$ with two commuting Killing isometries, the space of classical solutions is remarkably well understood because it can be constructed and analysed by algebraic methods \cite{Geroch:1972yt}. This does not come incidentally, but it is rather profoundly connected to the existence of hidden, infinite dimensional symmetries, that manifest themselves \cite{Geroch:1970nt} upon dimensionally reducing à la Kaluza-Klein (KK) the theory along the two isometries down to $D=2$. After the effective elimination of all the dependence on the additional coordinates, the Lagrangian of the $D=2$ theory can be cast in the form of a non-linear sigma model coupled to a dilaton and to $D=2$ gravity. This model is endowed with a hidden on-shell, infinite-dimensional duality symmetry of the Kac-Moody type, $\widehat{\text{SL}(2)}$ known as the Geroch group~\cite{Geroch:1972yt,Geroch:1970nt}, namely the central affine loop extension of $\text{SL}(2)$ (the underlying group theoretic structure thereof was first suggested by Julia \cite{Julia:1981wc, Julia:1980gr} and then elucidated in~\cite{Breitenlohner:1986um}). The action of the elements of the Geroch group maps solutions of the $D=2$ system into other solutions, its presence being ultimately the reason for the field equations' integrability, proven in~\cite{Belinsky:1971nt,Maison:1978es}.

A great body of literature has been devoted to the exploration of this two-dimensional system, in particular to its quantisation~\cite{Korotkin:1994au, Korotkin:1996vi, Korotkin:1996au, Korotkin:1996fx}, and to its generalisations to supergravity theories~\cite{Nicolai:1987kz,Nicolai:1988jb} and to exceptional field theory~\cite{Bossard:2018utw, Bossard:2021jix, Bossard:2023wgg}. Recently in~\cite{Cole:2024skp}, $D=2$ dimensionally reduced gravity has also been embedded into the framework of $D=4$ Chern-Simons theory~\cite{Costello:2017dso}, an approach that seems to describe most integrable field theories in $D=2$. Despite its rich mathematical structure, even small modifications of the system, like the addition of a cosmological constant $\Lambda$ in $D=4$ will generically spoil the integrability of the dimensionally reduced equations of motion: upon reduction to $D=2$, the different dilaton scaling power of $\Lambda$ would break the Geroch invariance, and the field equations for the dilaton and the conformal factor would not be separable anymore (some comments in this regard appear in \cite{Leigh:2014dja}). Nevertheless, it would be interesting for physical applications to allow for the presence of additional terms, like a dilaton potential $V(\rho)$ or like those arising in model building from gauged supergravities \cite{Ortiz:2012ib}, while preserving integrability and its solution generating techniques (for $D=1$ some reductions of this kind have been found  in~\cite{LopesCardoso:2024tol}). It is therefore a question of the utmost interest to determine how and to which extent such a theory can be deformed while preserving its integrability properties also for the $D=2$ case. 

Integrable $D=2$ field theories, of which the model just described represents a specific instance, are especially appealing, as the availability of analytic techniques in these contexts provides the possibility to obtain exact solutions to the equations of motion~\cite{Driezen:2021cpd}. Despite their interest, the occurrence of integrable field theories is, on the other hand, quite uncommon, and the search for new integrable models represents an active research area. In particular, great attention has been devoted in the literature to non-linear $\sigma$-models and on deformations thereof, e.g. Yang-Baxter deformations \cite{Klimcik:2002zj,Klimcik:2008eq,Delduc:2013qra}, $\lambda$-\cite{Sfetsos:2013wia} and WZ-term \cite{Klimcik:2017ken} deformations and their interplay \cite{Delduc:2014uaa}, or generalisations to non-symmetric cosets \cite{Young:2005jv,Hoare:2021dix,Osten:2021opf} and to the heterotic string \cite{Osten:2023cza}. For more details and references, see the recent reviews \cite{Driezen:2021cpd,Hoare:2021dix}.

A first analysis of how these integrable deformations might be combined with $D=2$ dimensionally reduced gravity appeared in \cite{Hoare:2020fye}. There, the authors consider several of the above integrable models but with `running' (space-time dependent) couplings, and a remarkable connection to the RG-flow of the couplings was observed. 

Recently, a new class of integrable deformations inspired by reformulations of Max\-well's theory~\cite{Ivanov:2001ec,Ivanov:2002ab,Ferko:2023wyi} has been introduced for Principal Chiral Models~\cite{Ferko:2024ali}. This approach relies on the introduction of auxiliary fields and of an arbitrary function of an invariant combination of such fields deforming the Lagrangian. Physically, these deformations correspond to the closed form of flows of theories generated by functions of $T\Bar{T}$ with $T$ being the energy-momentum tensor, generalising the $T\bar{T}$- and root-$T\bar{T}$-\linebreak deformations~\cite{Zamolodchikov:2004ce,Cavaglia:2016oda,Babaei-Aghbolagh:2022uij,Ferko:2022cix,Borsato:2022tmu}. These deformations, embedded into the framework of $D=4$ Chern–Simons theory \cite{Costello:2017dso} in \cite{Fukushima:2024nxm}, have been subsequently extended to higher-spin fields \cite{Bielli:2024ach}, Yang-Baxter deformations \cite{Bielli:2024fnp}, and to coset sigma models \cite{Cesaro:2024ipq,Bielli:2024oif}.

In this work, we construct new $D=2$ integrable field theories obtained as deformations of dimensionally reduced $D=4$ General Relativity, thereby providing new interesting examples of integrable models enriched by the presence of gravitational interactions. The focus of the work, in particular, lies on two structurally different families of deformations: the first one is obtained by further extending the auxiliary field approach of \cite{Ferko:2024ali} to the $D=2$ sigma model coupled to a dilaton and to two-dimensional gravity; the second one is attained by means of a Yang-Baxter deformation of the $D=2$ dilaton-gravity-matter coupling part of the Lagrangian \cite{Hoare:2020fye}, consistently matched with an appropriate deformation of the dilaton-gravity coupling term. This represents an extension of the approach in \cite{Hoare:2020fye} to Yang-Baxter deformations of coset $\sigma$-models, and a completion into a deformation of $D=2$ gravity.

The structure of the paper is the following. In section \ref{sec:Review}, we review the construction of the KK reduction of $D=4$ General Relativity down to $D=2$ discussed in the introduction. In particular, after setting the stage for the introduction of the relevant fields and their dynamics, we review the emergence of the Geroch duality group $\widehat{\text{SL(2)}}$ and of the integrability of the model. Section \ref{sec:AF} is devoted to the analysis of the Auxiliary Field Deformation of this model. We build the associated Lagrangian, provide a flat Lax connection, and prove the Hamiltonian integrability of the theory. In section \ref{sec:YB}, we construct the Yang-Baxter deformation of $D=2$ dimensionally reduced gravity, and furnish a flat Lax representation proving its Lax integrability. We conclude in section \ref{sec:Outlook} with some additional remarks and interesting future research directions. Some details about the group theoretic structures, the more technical computations, and the Yang-Baxter deformation of the Principal Chiral Model coupled to $D=2$ dimensionally reduced gravity are situated in the appendices. 

\section{Dimensionally reduced gravity}
\label{sec:Review}
The class of models we will focus on can be seen as arising from the dimensional reduction of $D=4$ General Relativity along two commuting space-like Killing isometries.
Here, we briefly review this setup and the emergence of $\widehat{\text{SL}(2)}$, partly following the notation of \cite{Katsimpouri:2015nqc} (for comprehensive treatments, see, for example, \cite{Nicolai:1991tt,Nicolai:1996pd}). We shall first introduce some details on the general structure of $\mathbb{Z}_2$ coset geometries, instrumental in the following construction.

Given a symmetric space $G/K$, the global isometry $G$ has a Lie algebra $\mathfrak{g}$ that decomposes as
\begin{align}
\label{eq:kp}
\mathfrak{g}= \mathfrak{g}^{(0)} \oplus \mathfrak{g}^{(1)} = \mathfrak{k} \oplus  \mathfrak{p}\, .
\end{align}
This decomposition is a $\mathbb{Z}_2$-grading of $\mathfrak{g}$ and $\mathfrak{k}$ is the Lie algebra of a (connected) subgroup $K\subset G$, fixed by an involution. 

The coset $\sigma$-model depends on a field $\mathbb{R}^{1,1} \to G/K$, where we refer to $\mathbb{R}^{1,1}$ as (flat) space-time.
We can choose a coset representative $\mathcal{V}(x)\in G$ 
that transforms as
\begin{align}
\label{eq:costrm}
\mathcal{V}(x) \to \mathcal{V}^\prime(x) =g^{-1}\mathcal{V}(x) k(x)\, ,
\end{align}
with global $g\in G$ and local (in space-time) $k(x) \in K$, corresponding to the freedom of choosing a coset representative of $G/K$ at each space-time point. Fixing a specific form of a representative for $G/K$  determines $k(x)$ as a compensating transformation that depends on $g$ and $\mathcal{V}(x)$. In the following and in the main text we have mostly suppressed the dependence on the space-time coordinate $x^\mu$ to avoid cluttering.

The  $\mathfrak{g}$-valued Maurer--Cartan current decomposes  according to~\eqref{eq:kp} as
\begin{align}
\label{eq:QP}
\mathcal{V}^{-1}\partial_\mu  \mathcal{V}=P_\mu +Q_\mu\,, \quad P_\mu\in\mathfrak{p}\, ,\; Q_\mu\in\mathfrak{k} .
\end{align}
The two components transform covariantly and as a connection, respectively, under~\eqref{eq:costrm}:
\begin{align}
\label{eq:QPtrm}
P_\mu \to  k^{-1}P_\mu k\,, \quad
Q_\mu  \to k^{-1}Q_\mu k+ k^{-1}\partial_\mu k\,.
\end{align}
From the definition~\eqref{eq:QP} one can deduce the Bianchi identities
\begin{align}
\label{eq:Bianchis}
2\partial_{[\mu}Q_{\nu]}+[Q_\mu,Q_\nu]=-[P_\mu,P_\nu]\, , \hspace{15mm} 
D_{[\mu}P_{\nu]}=0\, ,
\end{align}
where we have introduced the $K$-covariant derivative 
\be\label{eq:covder}
D_\mu = \partial_\mu +Q_\mu.
\ee
Armed with these definitions, we can now review the KK reduction to $D=2$ of $D=4$ General Relativity. 
We work with conventions $\eta_{AB}=\text{diag}(-,+,+,+)$, and, when useful, adopt lightcone coordinates in two dimensions,
\be\label{eq:lightonecoords}
x_{\pm}=\frac{1}{2}(t\pm x)\, .
\ee
Starting with the $D=4$ Einstein-Hilbert Lagrangian,
\begin{equation}\label{eq:EH}
 \mathcal{L}^{(4)}=\frac{1}{2}E_{(4)}R(E)_{(4)}\, ,   
\end{equation}
with $E_{(4)}$ the vierbein determinant and $R(E)_{(4)}$ the Ricci scalar, the aforementioned KK reduction can always be performed in (at least) two different ways: either by reducing directly on the two-torus $T^2$ parameterised by the two isometries, or by reducing first along one direction to $D=3$, dualising the KK vector to a scalar, and then finally reducing along the leftover isometry.

The first possibility is implemented through the choice of vierbein 
\begin{equation}\label{eq:vierbein1}
 E_M{}^A=\begin{pmatrix}
    f_{(\text{MM})} e_\mu{}^{\alpha} & B_\mu{}^m e_m{}^a\\
     0 & e_m{}^a
 \end{pmatrix},   
\end{equation}
where $\mu,\alpha=0,1$, $m, a=2,3$, $f_{(\text{MM})}$ is a conformal factor (with label to be clarified shortly), and $e_\mu{}^{\alpha}$ is the zweibein encoding the unimodular metric degrees of freedom $\eta_{\alpha\beta}e_\mu{}^{\alpha}e_\nu{}^{\beta}=h_{\mu\nu}$, with $\text{det}\,(h)=1$. The Kaluza-Klein vector $B_\mu{}^m$ carries no physical degree of freedom anymore in $D=2$, and acts as an auxiliary field (thus in this context to be effectively dropped). The lower-right block can be parameterised by its own determinant (measuring the size of $T^2$), $\rho\equiv \text{det} e_m{}^a$, and an SL$(2,\mathbb{R})$ matrix $\mathcal{V}_m{}^a$. Inserting directly \eqref{eq:vierbein1} (with $B_\mu{}^m=0$) into \eqref{eq:EH}, leads to the two-dimensional Lagrangian
\begin{equation}\label{eq:LagMM}
  \mathcal{L}^{(2)}{}_{(\text{MM})}= \frac{1}{2}e_{(2)}\rho R_{(2)}-\frac{1}{2}e_{(2)} \rho \text{Tr}(\tilde{P}_\mu \tilde{P}^\mu)+ e_{(2)} h^{\mu\nu}f^{-1}_{(\text{MM})}\partial_\mu f_{(\text{MM})}\partial_\nu\rho\, ,
\end{equation}
where $\tilde{P}_\mu$ is the $\text{SL}(2,\mathbb{R})/\text{SO}(2)$  coset current as defined in \eqref{eq:QP}.

The second option is instead achieved through an initial vierbein parameterisation
\begin{equation}\label{eq:vierbein2}
 E_M{}^A=\begin{pmatrix}
     \Delta^{-1/2}e_{\textsf{m}}{}^{\textsf{a}} & A_{\textsf{m}}\Delta^{1/2}\\
     0 & \Delta^{1/2}
 \end{pmatrix},   
\end{equation}
where $\mathsf{m},\mathsf{a}=0,1,2$, $\Delta$ is a scalar and $A_\mathsf{m}$ the KK vector in $D=3$.
Upon insertion of \eqref{eq:vierbein2} into \eqref{eq:EH}, we obtain
\begin{equation}
 \mathcal{L}_{(3)}=\frac{1}{2}e_{(3)}R_{(3)}-\frac{1}{4} e_{(3)} g^{\mathsf{m}\mathsf{n}}\Delta^{-1}\partial_\mathsf{m}\Delta\Delta^{-1}\partial_\mathsf{n}\Delta +\frac{1}{8}e_{(3)}\Delta^2 g^{\mathsf{p}\mathsf{m}}g^{\mathsf{n}\mathsf{q}}F_{\mathsf{m}\mathsf{n}} F_{\mathsf{p}\mathsf{q}},
\end{equation}
where $F_{\mathsf{m}\mathsf{n}}=\partial_{\mathsf{m}}A_{\mathsf{n}}-\partial_{\mathsf{n}}A_{\mathsf{m}}$. Dualising the field strength as
\begin{equation}
  \Delta^2 F_{\mathsf{m}\mathsf{n}}=\epsilon_{{\mathsf{m}\mathsf{n}}\mathsf{p}}\partial^{\mathsf{p}}B,  
\end{equation}
the three-dimensional Lagrangian can be rewritten as
\begin{equation}
\mathcal{L}_{(3)}=\frac{1}{2}e_{(3)}R_{(3)}- \frac{1}{2}e_{(3)}\text{Tr}(P_{\mathsf{m}} P^{\mathsf{m}} )\, ,
\end{equation}
where 
\begin{equation}
  P_{\mathsf{m}} =\begin{pmatrix}
      \frac{1}{2}\Delta^{-1}\partial_\mathsf{m}\Delta & \frac{1}{2}\Delta^{-1}\partial_\mathsf{m}B\\
      0 & -\frac{1}{2}\Delta^{-1}\partial_\mathsf{m}\Delta
  \end{pmatrix}\, 
\end{equation}
is the $\text{SL}(2,\mathbb{R})/\text{SO}(2)$ coset current, as defined in \eqref{eq:QP}.
The additional reduction along the remaining isometry direction is performed through the further parametrisation
\begin{equation}
  e_{\mathsf{m}}{}^a=\begin{pmatrix}
      f_{(\text{E})} e_\mu{}^{\alpha} & 0\\
      0 & \rho
  \end{pmatrix},  
\end{equation}
leading to 
\begin{equation}\label{eq:LagE}
    \mathcal{L}^{(2)}{}_{(\text{E})}=\frac{1}{2}e_{(2)}\rho R_{(2)}-\frac{1}{2}e_{(2)}\rho \text{Tr}(P_\mu P^\mu)+e_{(2)} h^{\mu\nu}f_{(\text{E})}^{-1}\partial_\mu f_{(\text{E})}\partial_\nu\rho\,\, .
\end{equation}

Two comments are in order in comparing \eqref{eq:LagMM} and \eqref{eq:LagE}. First, despite being both $D=2$ gravities coupled to an SL$(2)$ coset sigma model, their respective SL$(2)$'s are different (whence the tilded notation for the currents in \eqref{eq:LagMM}): in the first case the group is known as the \textit{Matzner-Misner} SL$(2)$, whereas in the second case the SL$(2)$ emerges already at the $D=3$ level after the vector dualisation, and it goes by the name of the \textit{Ehlers} SL$(2)$. This also explains the labels of the conformal factors. As the two theories are on-shell equivalent and the two groups act on scalars on-shell dual to each other \cite{Breitenlohner:1986um}, their consistent, simultaneous action can only be implemented via a symmetry enhancement to the central affine loop extension of SL$(2)$, the Geroch group $\widehat{\text{SL}(2)}$  \cite{Geroch:1972yt,Geroch:1970nt} (some details on the algebra of which are provided in appendix \ref{app:cosetmodels}). This enhancement of the on-shell symmetry to an infinite dimensional one can be more explicitly seen at the level of the equations of motion, with the occurrence of an infinite tower of dual potentials encoding the on-shell degrees of freedom and whose integrability conditions are equivalent to the field equations.

Secondly, the two Lagrangians can be further simplified by choosing the conformal gauge for the $D=2$ unimodular metric, namely
\begin{equation}
 h_{\mu\nu}= e^{2 \sigma} \eta_{\mu\nu}.  
\end{equation}
The factors $f_{(\text{E})}$ and $f_{(\text{MM})}$ can be reabsorbed into $e^{ \sigma}$, leading, respectively, to
\begin{align}\label{eq:LagMMgconf}
&\mathcal{L}^{(2)}{}_{(\text{MM})}= \partial_\mu\rho\partial^\mu\tilde{\sigma}- \frac{1}{2}\rho \text{Tr}(\tilde{P}_\mu \tilde{P}^\mu)\,,\\
 &\mathcal{L}^{(2)}{}_{(\text{E})}= \partial_\mu\rho\partial^\mu\sigma- \frac{1}{2}\rho \text{Tr}(P_\mu P^\mu)\,  ,\label{eq:LagEgconf} 
\end{align}
where tildes denote the Matzner-Misner scalar fields, to remark their difference with the Ehlers ones. The Lagrangians \eqref{eq:LagMMgconf}, \eqref{eq:LagEgconf} must be supplemented  with their associated Virasoro constraints: the latter stem from a combination of the diffeomorphism and the Hamiltonian constraints,  formally obtained by varying the actions with respect to the unimodular degrees of freedom before gauge-fixing. In lightcone coordinates \eqref
{eq:lightonecoords}, the Virasoro constraints read
\begin{align}
&V_{(\text{MM})\; \pm\pm}=\partial_{\pm}\rho\partial_{\pm}\tilde{\sigma}-\frac{1}{2}\partial_{\pm}\partial_{\pm}\rho- \frac{1}{2}\rho \text{Tr}(\tilde{P}_{\pm} \tilde{P}_{\pm}) =0\, , \label{eq:Vir1}\\
&V_{(\text{E})\; \pm\pm}=\partial_{\pm}\rho\partial_{\pm}\sigma-\frac{1}{2}\partial_{\pm}\partial_{\pm}\rho-\frac{1}{2}\rho \text{Tr}(P_{\pm} P_{\pm})=0\,.\label{eq:Vir2}    
\end{align}

In the rest of the work, we will take the starting $D=2$ theory to be the Ehlers Lagrangian \eqref{eq:LagEgconf} with Virasoro constraint \eqref{eq:Vir2}, and the Geroch group encoding at the same time the on-shell degrees of freedom of the (on-shell) equivalent theory \eqref{eq:LagMMgconf}\footnote{An equivalent treatment could be developed starting from \eqref{eq:LagMMgconf}, albeit the action of the Geroch group on the physical fields would be more cumbersome to express.}.  The Euler-Lagrange equations stemming from \eqref{eq:LagEgconf} are \footnote{Notice that  equation \eqref{eq:boxsigma0} is not fundamental, and it is automatically implied if \eqref{eq:DP0},\eqref{eq:boxrho0}, \eqref{eq:Vir2} and \eqref{eq:Bianchis} are satisfied. Also notice that it can be equivalently formulated in terms of $\hat{\sigma}$ later defined in \eqref{eq:shiftedsigma}, as \textit{on-shell} $\Box\sigma\approx \Box \hat{\sigma}$. }
\begin{align}
 &D_\mu(\rho P^\mu)=0\,, \label{eq:DP0}\\
 &\Box\sigma+\frac{1}{2} \text{Tr}(P_\mu P^\mu)=0\,,\label{eq:boxsigma0}\\
 &\Box \rho=0\,\label{eq:boxrho0},
\end{align}
 together with the Bianchi identities \eqref{eq:Bianchis}. 

We remark here that the gravity-dilaton-sigma model structure \eqref{eq:LagEgconf}, \eqref{eq:Vir2} represents a general feature of $D=2$ KK reductions along commuting isometries not only of General Relativity from $D=4$ but also of General Relativity and of the bosonic sector of supergravities from $D\geq 4$ (the difference residing in the coset $\text{SL}(2)/ \text{SO}(2)$  replaced by a more general one $G/K$).
\subsection{Lax integrability}
\label{sec:Laxintegrability}
\subsubsection{Lax connection and the linear system}
Lax integrability of \eqref{eq:LagEgconf} can be shown by proving the existence of a Lax connection~\cite{Breitenlohner:1986um,Pohlmeyer:1975nb},
\begin{equation}\label{eq:Lax00}
 \mathfrak{L}_\mu(\gamma)=Q_\mu+\frac{1+\gamma^2}{1-\gamma^2}P_\mu+ \frac{2\gamma}{1-\gamma^2}\epsilon_{\mu\nu}P^\nu\, ,  
\end{equation}
whose flatness is equivalent to the Euler-Lagrange equations \eqref{eq:DP0}, 
\eqref{eq:boxrho0}. Here, $\gamma=\gamma(t,x; z)$ is a spacetime-dependent spectral parameter, itself a function of the constant spectral parameter $z$, and must satisfy
\begin{equation}
    \label{eq:linsystemrho}
    \frac{\partial_\mu\gamma}{\gamma}=\frac{1+\gamma^2}{1-\gamma^2}\frac{\partial_\mu\rho}{\rho}+\frac{2 \gamma}{1-\gamma^2}\epsilon_{\mu\nu}\frac{\partial^\nu\rho}{\rho}\, .
\end{equation}
Notice that equation \eqref{eq:boxrho0} allows for the definition of a dual partner of the dilaton,
\begin{equation}
    \label{eq:tilderho}
\partial_\mu\rho=\epsilon_{\mu\nu}\partial^\nu\tilde{\rho}\,,
\end{equation}
whose integrability condition is equivalent to equation \eqref{eq:boxrho0}. With such definition in mind, \eqref{eq:linsystemrho} is solved in particular by
\begin{equation}
    \label{eq:gamma}
    \gamma(\rho,\tilde{\rho};\,z)=\frac{1}{\rho}\left(z+\tilde{\rho}-\sqrt{(z+\tilde{\rho})^2-\rho^2}\right)\, ,
\end{equation}
thereby displaying that the spacetime dependence of $\gamma$ is implicit through $\rho$ and $\tilde{\rho}$.  The connection \eqref{eq:Lax00} bears no information regarding equation \eqref{eq:Vir2}, treated in these regards as a constraint to be imposed on the dynamics.

An equivalent statement of the integrability of the model is given in terms of the linear system \cite{Breitenlohner:1986um,Belinsky:1971nt,Maison:1978es}, which relies on an explicit parametrisation of the $\widehat{\text{SL}(2)}$ coset representative $\hat{\mathcal{V}}(z)$, in such a way that 
\be\label{eq:linearsystem0}
\hat{\mathcal{V}}^{-1}(z)\partial_\mu\hat{\mathcal{V}}(z)= \mathfrak{L}_\mu(\gamma)\,,
\ee
with $\mathfrak{L}_\mu(\gamma)$ defined as in \eqref{eq:Lax00}. Flatness of the Lax connection \eqref{eq:Lax00} then is tantamount to the compatibility of the linear system \eqref{eq:linearsystem0}.
Remarkably, a more general linear system highlighting the presence of an even larger group-theoretic structure underlying the model can be constructed explicitly \cite{Julia:1996nu,Bossard:2021jix} (see also \cite{Paulot:2004hh}). This realises the presence of a coset-like symmetry,
\be\label{eq:largercoset}
\frac{\mathcal{W}\ltimes \widehat{\text{SL}(2)}}{K\left(\mathcal{W}\right)\ltimes K\left(\widehat{\text{SL}(2)}\right)},
\ee
whose first glimpses were observed in \cite{Maison:1988zx} and  \cite{Hou:1987xr}, and which can be made manifest by the introduction of an infinite set of supplementary fields, as to be shown shortly. In \eqref{eq:largercoset}, $K\left(\widehat{\text{SL}(2)}\right)$ is the maximal compact subgroup of $\widehat{\text{SL}(2)}$, while the definition of  $\mathcal{W}$ and $K(\mathcal{W})$ is more subtle \cite{Julia:1996nu}. $\mathcal{W}$, which we will call the Witt-Virasoro  ``group'', should be thought of as the group of real analytic
reparametrisations $z \rightarrow f(z)$ of the constant spectral parameter $z$ (although one needs to be careful with its precise definition \cite{Julia:1996nu}), while $K(\mathcal{W})$ plays the role of its ``maximal compact subgroup", and can be thought of as the real group of transformations of the spacetime dependent spectral parameter $\gamma$ generated  by the operators ($m\geq1$)
\begin{equation}
\mathsf{k}_m:=-(\gamma^{m+1}-\gamma^{-m+1})\frac{\partial}{\partial\gamma}.
\end{equation}

The approach relies on the introduction of a formal element of the Witt-Virasoro group (in the triangular gauge) $\mathbf{\Gamma}\in\mathcal{W}$ \cite{Julia:1996nu}, \cite{Bossard:2021jix}, built out of an infinite set of ancillary fields $\phi_m(x)$ with $m\geq 1$,
\be
\mathbf{\Gamma}(x)=... e^{-\phi_2(x) L_2}e^{-\phi_1(x) L_1}\rho^{-L_0}\, ,
\ee 
where the Witt-Virasoro generators $L_m$ are defined in \eqref{eq:Virgens}.
The $\widehat{\text{SL}(2)}$ coset representative $\hat{\mathcal{V}}$ can then be extended to its generalisation
\be\label{eq:genV}
\hat{\mathcal{V}}\, \rightarrow \hat{\mathcal{Y}}=\left(\hat{\mathcal{V}}(z)\cdot\mathbf{\Gamma},\,e^{\hat{\sigma}}\,\mathsf{K}\right)\in\frac{\mathcal{W}\ltimes \widehat{\text{SL}(2)}}{K\left(\mathcal{W}\right)\ltimes K\left(\widehat{\text{SL}(2)}\right)}\, ,
\ee
where the second entry in \eqref{eq:genV} features the presence of the shifted conformal factor \footnote{A covariant analog definition of \eqref{eq:shiftedsigma} that does not rely on the conformal gauge in lightcone coordinates can be given by means of Beltrami differentials \cite{Nicolai:1993pe}.}
\be\label{eq:shiftedsigma}
\hat{\sigma}\equiv \sigma-\frac{1}{2}\text{ln}\left(\partial_+\rho\partial_-\rho\right)\, ,
\ee
and is built in order to accommodate for the presence of the central charge. In the parenthesis \eqref{eq:genV} we have let the $\widehat{\text{SL}(2)}$ central factor $\mathsf{K}$ stand out, as it will be important in what follows. 

The enlarged linear system reads
\begin{align}\label{eq:largeLS}
\nonumber \hat{\mathcal{Y}}^{-1}\partial_\mu\hat{\mathcal{Y}}&= \left(\mathbf{\Gamma}^{-1}\hat{\mathcal{V}}^{-1}\partial_\mu\left(\hat{\mathcal{V}}\mathbf{\Gamma}\right), \, \left[\partial_\mu \hat{\sigma}-\Omega^{\prime}(\hat{\mathcal{V}}, \hat{\mathcal{V}}^{-1}\partial_\mu \hat{\mathcal{V}})\right]\mathsf{K}\right)\cong(\mathfrak{L}_{\mu}(z),0)\\[3pt]
 &\equiv\left(Q_\mu+\frac{1+z^2}{1-z^2}P_\mu+\frac{2 z}{1-z^2}\epsilon_{\mu\nu}P^\nu+\left(\frac{1+z^2}{1-z^2}\frac{\partial_\mu\rho}{\rho}+\frac{2 z}{1-z^2}\epsilon_{\mu\nu}\frac{\partial^\nu\rho}{\rho}\right)z\frac{\partial}{\partial z},\,0\right),
\end{align}
where $\Omega^{\prime}(\hat{\mathcal{V}}, \hat{\mathcal{V}}^{-1}\partial_\mu \hat{\mathcal{V}})$ is the mixed cocycle introduced in \cite{Breitenlohner:1986um}. The symbol $\cong$ denotes equality upon imposition of a twisted self-duality constraint \cite{Julia:1996nu}; this constraint guarantees the recovery of the connection displayed in the second line and, relating the infinite tower $\phi_m(x)$, for $m > 1$, to $\phi_1(x)=\tilde{\rho}$ and $\rho$ via duality relations, singles out the correct number of degrees of freedom.  Notice that the action of $\mathbf{\Gamma}$ allowed for the replacement of the spacetime dependent spectral parameter $\gamma(\rho,\tilde{\rho};\,z)$ by the constant one $z$, at the expense of introducing explicitly the linear system for $\rho$ by virtue of $\mathbf{\Gamma}^{-1}\partial_\mu\mathbf{\Gamma}$ \cite{Bossard:2021jix}.

Furthermore, the last entry in the second line is now null, as demanded by the requirement of invariance under the involutory automorphism $\tau^{\infty}$ of the linear system \cite{Julia:1996nu}. This amounts to the vanishing of the $\mathsf{K}$ coefficient in the first line, which returns a first order equation for $\sigma$; an explicit computation \cite{Breitenlohner:1986um}, \cite{Paulot:2004hh}, shows that this is precisely the Virasoro constraint \eqref{eq:Vir2}. Flatness of \eqref{eq:largeLS} is then equivalent to \eqref{eq:DP0} and \eqref{eq:boxrho0}, with \eqref{eq:boxsigma0} following from the Virasoro constraint. 
 
\subsubsection{A centrally extended Lax connection }\label{subsec:Lax}
An alternative way to reproduce all the equations of motion \eqref{eq:DP0}, \eqref{eq:boxsigma0}, \eqref{eq:boxrho0} is the following extended Lax connection, valued in $\hat{\mathfrak{g}} \rtimes \text{Vir}$ (whose form had been suggested already in \cite{Bernard:2000hb,Bernard:2001pp}),
\begin{equation}
    \mathfrak{L}_{\text{ext}\, \mu} = Q_\mu + \frac{1+z^2}{1-z^2} P_\mu + \frac{2z}{1-z^2}  \epsilon_{\mu\nu} P^\nu + \left( \frac{1+z^2}{1-z^2} \frac{\partial_\mu \rho}{\rho} + \frac{2z}{1-z^2} \epsilon_{\mu \nu} \frac{  \partial^\nu \rho}{\rho} \right) L_0 + \epsilon_{\mu\nu} \partial^\nu\sigma\, \mathsf{K} \, ,\label{eq:UndefLax}
\end{equation}
with $L_0=-z\frac{\partial}{\partial z}$, and where we have added a central charge term. With this formalism, it turns convenient to work in a different parametrisation of the spectral parameter $z$,
\begin{align}\label{eq:spreparam}
    z = \frac{1-\lambda}{1+\lambda}\, ,
\end{align} 
and with a set of new, now Virasoro,  algebra generators $\mathsf{L}_m$ realised in terms of the latter parameter as
\begin{align}
    \mathsf{L}_m = -\frac{1}{2} \lambda^{2m+1} \frac{\partial}{\partial \lambda}\,,
\end{align}
 and satisfying 
 \begin{equation}
    \label{eq:KVir}
 [\mathsf{L}_m, \mathsf{L}_n]=(m-n)\mathsf{L}_{m+n}+\frac{c_{\mathfrak{vir}}}{12}m(m-1)\delta_{m, -n}\mathsf{K}   \, .
\end{equation} 
In \eqref{eq:KVir}, $c_{\mathfrak{vir}}$ is a central charge determined by the weight of the representation upon which $\mathsf{L}_m$ acts, but we will never need it in the following. The loop generators $T_m^A$ are now realised as $T_m^A=T^A\lambda^m$, they close the algebra

 \begin{equation}
     [T^A_m , T^B_n ] = f^{AB}{}_C T^C_{m+n} + \eta^{AB} \frac{m}{2} \delta_{m+n,0} \mathsf{K}\, ,
 \end{equation} 
 and the semidirect action of the Virasoro algebra on the loop part is
 \begin{equation}
     [\mathsf{L}_m, T_n^A]=-\frac{n}{2}T^A_{2m+n}\, .
 \end{equation}
Let us remark that $\mathsf{L}_{m}$ \textit{cannot} be obtained from $L_m$ by applying the straightforward reparametrisation \eqref{eq:spreparam}. With these conventions, we are able to match the connection \eqref{eq:UndefLax} to the one appearing in \cite{Bernard:2000hb}. 

The functions of $z$ in \eqref{eq:UndefLax} must be now meant as series expansions in terms of $\lambda$, and, in particular
\begin{align}\label{eq:nicecomm}
  \nonumber  \left[ A_\mu \frac{1+z^2}{1 - z^2} , B_\nu \frac{2z}{1 - z^2} \right] &= \left[ A _\mu\,\frac{1}{2} \left( \frac{1}{\lambda} + \lambda \right) , B_\nu\,  \frac{1}{2}\left( \frac{1}{\lambda} - \lambda \right) \right]\\[2pt] 
    &    =[A_\mu,B_\nu]\, \frac{2z (1+z^2)}{(1 - z^2)^2} + \frac{1}{4} \text{Tr}(A_\mu B_\nu)\mathsf{K} \,,
\end{align}
 for $A_\mu, B_\nu\in \mathfrak{g}$, which is the only combination needed in the subsequent computations. The flatness of the Lax connection for all values of the constant spectral parameter $z$ (we adopt here the form notation in order to avoid cluttering)
\begin{align}
    0 &= \mathrm{d} \mathfrak{L}_{\text{ext}} + \frac{1}{2} [ \mathfrak{L}_{\text{ext}} , \mathfrak{L}_{\text{ext}} ] \\
    &= \left( \mathrm{d} Q + \frac{1}{2} [Q,Q] + \frac{1}{2} [P,P] \right) + \frac{1+z^2}{1 - z^2} \left(  \mathrm{d}P + [Q,P] \right) + \frac{2z}{1-z^2} \left( \mathrm{d} \star P + [Q, \star P] \right) \nonumber \\
    &{} \qquad + \frac{2z}{1-z^2} \left( \Box \rho \right) L_0 + \left( \Box \sigma + \frac{1}{2} \text{Tr} (P \wedge \star P) \right) \mathsf{K}\, , \nonumber
\end{align}
is equivalent to the equations of motion \eqref{eq:DP0}, \eqref{eq:boxrho0}, \eqref{eq:boxsigma0} for $\rho$, $\sigma$ and $\mathcal{V}$, and the Maurer-Cartan identity \eqref{eq:Bianchis} for $Q,P$. In order to reach this result, we have made use of \eqref{eq:nicecomm} and of the fact that there exists no invariant bilinear pairing between $L_0$ and $T_A$.

However, we here want to stress that, although corresponding to the field equations through its flatness, not only is the Lax connection \eqref{eq:UndefLax} oblivious to the Virasoro constraints \eqref{eq:Vir2}, but, due to the presence of the central charge, it is not even invariant under $\tau^\infty$ anymore. Therefore, the extent of its usefulness in the construction of the monodromy matrix, which is usually spacetime independent precisely because of such invariance, remains unclear. On the other hand, the present underlying structure strongly resembles the one appearing in Affine Gaudin models \cite{Lacroix:2018itd,Lacroix:2023gig} -- due to appearance of Kac-Moody algebras and central extensions, but not one related to the spectral parameter.\footnote{We thank Sylvain Lacroix for pointing this out to us.}

\subsection{Hamiltonian integrability}
The Hamiltonian integrability of \eqref{eq:LagEgconf}, namely the existence of infinitely many, linearly independent, Poisson commuting conserved charges, has been explored in different contexts (see for example \cite{Bernard:2000hb,Bernard:2001pp,Korotkin:1997fi,Samtleben:1998fk}). One way to achieve this is to show that the Poisson brackets of the spatial components of the Lax connection take a specific form, a result proven by Maillet in \cite{Maillet:1985ec,Maillet:1986}. In \cite{Korotkin:1997fi}, this was shown to be the case for the Lax connection \eqref{eq:Lax00}, whose spectral parameter $\gamma$ is spacetime dependent, at least on the constraint surface parameterising the physical phase space. Alternatively, \cite{Bernard:2001pp} reached the same conclusion for the connection \eqref{eq:UndefLax}, in terms of the constant spectral parameter $\lambda$, on the full phase space: their results partially match the ones in \cite{Korotkin:1997fi}, at least where comparison is possible, namely for the $r$-matrix part related to the coset model on the constraint surface.


Given canonical variables $\{\mathcal{V}, \rho, \sigma\}$ the canonical momenta obtained from \eqref{eq:LagEgconf} are 
\begin{align}
    \pi^{(\mathcal{V})} &= \rho P_t \mathcal{V}^{-1}\, , \qquad \pi^{(\rho)} =- \partial_t \sigma\, , \qquad \pi^{(\sigma)}= -\partial_t \rho.
\end{align}
The associated non-vanishing canonical Poisson brackets read
\begin{align*}
    \{ \pi^{(\rho)} (x) , \rho (y) \} ) &=\{ \pi^{(\sigma)} (x) , \sigma(y) \} = - \delta(x-y),  \\
    \{ \pi_{\mathbf{1}}^{(P)}(x) , \pi_{\mathbf{2}}^{(P)}(y) \} &= [ \Omega_{\mathfrak{p}} , \pi^{(Q)}_{\mathbf{2}}] \delta(x-y) \approx 0,\\
    \{ \pi_{\mathbf{1}}^{(P)}(x) , \pi_{\mathbf{2}}^{(Q)}(y) \} &= [ \Omega_{\mathfrak{k}} , \pi^{(P)}_{\mathbf{2}}] \delta(x-y), \\
    \{ \pi_{\mathbf{1}}^{(Q)}(x) , \pi_{\mathbf{2}}^{(Q)}(y) \} &= [ \Omega_{\mathfrak{k}} , \pi^{(Q)}_{\mathbf{2}}] \delta(x-y) \approx 0, \\
    \{ \pi_{\mathbf{1}}^{(P)}(x) , P_{x,\mathbf{2}}(y) \} &= [ \Omega_{\mathfrak{p}} , Q_{x,\mathbf{2}}] \delta(x-y) + \Omega_{\mathfrak{p}} \partial_x \delta(x-y), \\
     \{ \pi_{\mathbf{1}}^{(Q)}(x) , Q_{x,\mathbf{2}}(y) \} &= [ \Omega_{\mathfrak{p}} , Q_{x,\mathbf{2}}] \delta(x-y) + \Omega_{\mathfrak{k}} \partial_x \delta(x-y).
\end{align*}
We introduced $\pi^{(Q)}$ as the canonical variable $\pi^{(Q)} \approx 0$ is the constraint that generates the $K$-gauge symmetry. We used the standard notation $V_{\mathbf{1}} = V \otimes 1$, $V_{\mathbf{2}} = 1 \otimes V$, where $\Omega = \Omega_{\mathfrak{k}} + \Omega_{\mathfrak{p}} = T_A \otimes T^A$ is the tensorial Casimir on $\mathfrak{g}$ decomposed into the $\mathbb{Z}_2$-grading $\mathfrak{g} = \mathfrak{k} \oplus \mathfrak{p}$.
In these canonical variables the Lax matrix $\mathfrak{L}_{ext,x}$ is
\begin{align}
    \mathfrak{L}_{ext,x}(x,z) &= Q_x + f(z) \pi^{(Q)} + \frac{1+z^2}{1-z^2} P_x + \frac{1}{ \rho} \frac{2z}{1-z^2} \pi^{(P)} \nonumber \\
    &{} \quad + \left( \frac{1+z^2}{1-z^2} \frac{\partial_x \rho}{\rho} - \frac{1}{\rho} \frac{2z}{1-z^2} \pi^{(\sigma)} \right) L_0 - \pi^{(\rho)} \mathsf{K}. \label{eq:ExtLaxUndeformedCanonical}
\end{align}
As in \cite{Lacroix:2018njs, Cesaro:2024ipq}, the addition of the term $f(z) \pi^{(Q)} \approx 0$ is used, s.t. the Poisson bracket $\lb \mathfrak{L}_{ext,x,\mathbf{1}}(x,z), \mathfrak{L}_{ext,x,\mathbf{2}}(x',z')\rb$ can be written Maillet form. Here,
\begin{equation}
    f(z) = - \frac{z}{(1+z)^2} \ .
\end{equation}
With the classical $r$-matrix\footnote{We remark that $L_0=-z\frac{\partial}{\partial z}$.}
\begin{equation}
    r_{\mathbf{12}} (z,z^\prime) = 
    \frac{(f(z^\prime))^2}{f(z^\prime) - f(z)} \left( \Omega_{\mathfrak{k}} + \frac{1-z}{1+z} \frac{1+z^\prime}{1-z^\prime} \Omega_{\mathfrak{p}} \right) + \frac{1 + z^2}{1 - z^2} L_0 \otimes \mathsf{K} \, ,\label{eq:ExtLaxRmatrix}
\end{equation}
it is possible to show that the Poisson bracket of the extended Lax connection
\begin{align}
\label{eq:Maillet}
    &{} \quad \lb \mathfrak{L}_{ext,x,\mathbf{1}}(x,z), \mathfrak{L}_{ext,x,\mathbf{2}}(x',z')\rb \nn \\
    &= 
    \frac{1}{\rho(x)} \left( \left[ r_{\mathbf{12}}(z,z^\prime),\mathfrak{L}_{x, \mathbf{1}}(x,z)\right]
    + \left[r_{\mathbf{21}}(z^\prime, z)\mathfrak{L}_{x, \mathbf{2}}(x',z^\prime)\right] \right) \delta(x-x^\prime) \\
&{} \quad - \frac{1}{2} \left( \frac{1}{\rho(x)} + \frac{1}{\rho(x^\prime)} \right) \left(r_{\mathbf{12}}(z,z^\prime)+r_{\mathbf{21}}(z^\prime, z)\right)\partial_{x}\delta(x{-}x')\, , \nn
\end{align}
fits into the Maillet bracket form \cite{Maillet:1985ec,Maillet:1986}. In contrast to \cite{Bernard:2000hb}, we chose to stick to the conventions associated to $z$, rather than to $\lambda$, in such a way as to display \eqref{eq:Maillet} in a more conventional form.

\section{Auxiliary Field Deformation}
\label{sec:AF}
In this section we turn to the first of the models of interest in this work. Its formulation relies on the deformation of the Lagrangian \eqref{eq:LagEgconf} through the introduction of auxiliary fields, an approach fruitfully pioneered in \cite{Ferko:2024ali} in order to deform the Principal Chiral Model while preserving its integrability, and subsequently extended in \cite{Bielli:2024ach,Bielli:2024fnp,Cesaro:2024ipq,Bielli:2024oif}. We provide a flat Lax representation for the system and a straightforward argument for the involution of its conserved charges, thereby establishing its complete integrability.
\subsection{The Lagrangian}
 A deformation of \eqref{eq:LagEgconf} by fields auxiliary to the coset structure alone turns out not to be integrable. On the grounds of the coupling (via the dilaton field) between the coset and the conformal factor -- dilaton system, this is to be expected. In order to obtain an integrable deformation through the auxiliary field approach, the kinetic term for the aforementioned system and the coset term must be treated on an equal footing, and accordingly deformed by the introduction of appropriate auxiliary fields. Therefore, we start from the ansatz 
\begin{align}\label{eq:Ldef}
\nonumber \tilde{\mathcal{L}}&=-\partial_\mu \rho\partial^\mu\sigma -2(\chi_{1\,\mu} \partial^\mu \rho+\chi_{2\,\mu}\partial^\mu\sigma+\chi_1^\mu\chi_{2\,\mu})\\
&{} \quad +\rho\left(\text{Tr}\left[\frac{1}{2}(P_\mu P^\mu)+2P_\mu v^\mu+v_\mu v^\mu\right]\right)+E(\nu)\, .
\end{align}
Here, $\left(\chi_{1\,\mu}, \chi_{2\,\mu}, v_\mu\right)$ are a set of auxiliary fields (i.e. their equations of motion are algebraic), with $v_\mu \in \mathfrak{p}$; $E(\nu)$ is an arbitrary function of the parameter $\nu$ ,
\begin{equation}\label{eq:nu}
\nu=(\eta^{\alpha\beta}\eta^{\gamma\sigma}+\epsilon^{\alpha\beta}\epsilon^{\gamma\sigma})\left(\chi_{1\,\alpha}\chi_{2\, \gamma}-\frac{\rho}{2}\text{Tr}(v_\alpha v_\gamma)\right)\left(\chi_{1\,\beta}\chi_{2\, \sigma}-\frac{\rho}{2}\text{Tr}(v_\beta v_\sigma)\right)\, .
\end{equation}
Defining 
\be\label{eq:J}
\mathcal{P}_\mu=-(P_\mu+2v_\mu)\, ,\quad \mathcal{R}_{\, \mu}=-(\partial_\mu\rho+2\chi_{2\,\mu})\, ,\quad \mathcal{S}_{\, \mu}=-(\partial_\mu\sigma+2\chi_{1\,\mu})\, ,
\ee
the corresponding Euler-Lagrange equations read 
\begin{align}
\partial_\mu\mathcal{R}_{}^\mu&=0\, ,\label{eq:Boxrho=smth}\\
D_\mu(\rho\mathcal{P}^\mu)&=0\, ,\label{eq:dynamicaleom} \\
\partial_\mu \mathcal{S}^\mu&=\text{Tr}\left[\frac{1}{2}(P_\mu P^\mu)+2P_\mu v^\mu+v_\mu v^\mu\right]+K^{\mu\nu}(\chi_{1,2},v,\rho)\,\text{Tr}(v_\mu v_\nu) \, ,\label{eq:Boxsigma+smth} \\
P_\mu&=-v_\mu+ K_\mu{}^\nu(\chi_{1,2},v,\rho) v_\nu,\label{eq:algebraicP}\\
\partial_\mu\sigma&=-\chi_{1\,\mu}+ K_\mu{}^\nu(\chi_{1,2},v,\rho)\chi_{1\,\nu}\, ,\label{eq:algebraicsigma}\\
\partial_\mu\rho &=-\chi_{2\,\mu}+K_\mu{}^\nu(\chi_{1,2},v,\rho)\chi_{2\,\nu}\, ,\label{eq:algebraicrho}
\end{align}
where
\begin{equation}
K_\mu{}^\nu(\chi_{1,2},v,\rho)=E^\prime(\nu)(\delta_\mu^\alpha\eta^{\gamma\nu}-\eta_{\mu\rho}\epsilon^{\rho\alpha}\epsilon^{\gamma\nu})\left(\chi_{1\,\alpha}\chi_{2\,\gamma}-\frac{\rho}{2}\text{Tr}(v_\alpha v_\gamma)\right)\,  
\end{equation}
is symmetric in its indices $(\mu,\nu)$ when both raised or lowered by the flat metric, and $E^{\prime}(\nu)=\frac{\partial E(\nu)}{\partial \nu}$. In passing, let us highlight the useful property
\begin{equation}\label{eq:epsK}
\epsilon^{\mu\nu}K_\mu{}^\sigma = -\epsilon^{\mu\sigma}K_\mu{}^\nu\, ,   
\end{equation}
which will turn out to be crucial in subsequent computations. For calculations it is often useful to use light-cone coordinates \eqref{eq:lightonecoords}. Then the tensor $K$ is specified as follows in terms of the deformation defining function $E(\nu)$:
\begin{align}
    {K_\pm}^{\mp} &= \frac{1}{2} E^\prime(\nu) \left( \chi_{1\pm} \chi_{2\pm} - \frac{1}{2} \rho \text{Tr} (v_\pm v_\pm) \right), \qquad {K_\pm}^{\pm} = 0. \label{eq:AFDefK}
\end{align}
In order to highlight how all equations of motion are deformed through the defining function $E$, respectively the tensor $K(v,\chi_1,\chi_2)$, using the algebraic equations of motion, the dynamical ones can be written in form notation as
\begin{align}
    D\left(\rho \star \frac{1+K}{1-K} P \right) &= 0, \label{eq:AFPdef} \\
    \mathrm{d} \left( \star \frac{1+K}{1-K} \mathrm{d} \rho \right) &= 0, \label{eq:AFRhodef} \\
    \mathrm{d} \left( \star \frac{1+K}{1-K} \mathrm{d} \sigma \right) &= - \frac{1}{2} \text{Tr} (P \wedge \star \mathcal{P}) = - \frac{1}{2} \text{Tr} \left( P \wedge \star \frac{1+K}{1-K} P \right), \label{eq:AFSigmadef}
\end{align}
where $v,\chi_1,\chi_2$ still implicitly depend on $P,\mathrm{d}\rho , \mathrm{d}\sigma$ through their respective algebraic equations of motion. 

\subsection{Auxiliary fields as deformations}
The model \eqref{eq:Ldef} is a smooth deformation of the Lagrangian \eqref{eq:LagEgconf}. Let us check the explicit form $E(\nu)=c F(\nu)$, for some small constant $c$. Then the Lagrangian \eqref{eq:Ldef} becomes:
\begin{align}
    \mathcal{L}_c &= \partial_\mu \rho \partial^\mu \sigma - \frac{\rho}{2} \text{Tr} \left( P_\mu P^\mu \right) \label{eq:LdefLin} \\
    &{} \quad + cF\left( \left( \partial_+ \rho \partial_+ \sigma - \frac{\rho}{2} \text{Tr}(P_+ P_+) \right) \left(\partial_- \rho \partial_- \sigma - \frac{\rho}{2} \text{Tr}(P_- P_-) \right) \right) + \mathcal{O}(c^2). \nonumber
\end{align}
at linear order. The auxiliary fields $v,\chi_1, \chi_2 $ can be integrated out order by order in $c$, e.g.:
\begin{equation}
    v = -\left(1+K(P,\partial_\mu \rho,\partial_\mu \sigma) \right) P + \mathcal{O}(c^2),
\end{equation}
where $K$ is linear in $c$. One recognises that \eqref{eq:LdefLin} is indeed a deformation of \eqref{eq:LagEgconf} by a function $F$. At linear order the argument of the defining function $F$ is of the form $T\Bar{T}$ with the \textit{naive} energy momentum tensor $t$
\begin{equation}
    t_{\pm \pm} = \partial_\pm \rho \partial_\pm \sigma - \frac{\rho}{2} \text{Tr}(P_\pm P_\pm).
\end{equation}
This fits into the pattern of the Auxiliary Field Deformations for $\sigma$-models as generalisations of $T\Bar{T}$- and root-$T\Bar{T}$-deformations \cite{Ferko:2024ali}. Let us note that this naive energy momentum tensor $t$ does \textit{not} coincide with the Virasoro constraints  \eqref{eq:Vir2} of the underlying gravity theory and, hence, is \textit{not} zero but instead, under the Virasoro constraints, $t_{\pm \pm} \approx \partial_\pm \partial_\pm \rho + \mathcal{O}(c)$. 

Another characteristic feature of such deformations is the intrinsic presence of higher derivative terms. In the simplest case, for example,  $F(\nu) = \nu$, this introduces quartic interactions at first order.

At linear order, the equations of motion \eqref{eq:AFPdef}, \eqref{eq:AFRhodef} and \eqref{eq:AFSigmadef} simplify and can be expressed only in terms of the original physical fields, i.e. without referring to the auxiliary fields. For example, the equation of motion for $\rho$ can be written as
\begin{align}
    0 &= \partial_+ \partial_- \rho + c \partial_+ (F^\prime (t_{++} t_{--}) t_{--} \partial_+ \rho) + c \partial_- ( F^\prime (t_{++} t_{--}) t_{++} \partial_- \rho) + \mathcal{O}(c^2).
\end{align}
Using the Virasoro constraints, this equation decouples from the others at linear order:
\begin{align}
        0 = \partial_+ \partial_- \rho   \label{eq:DilatonAUX} &+ c  \partial_+ \left(F^\prime (\partial_+ \partial_+ \rho \ \partial_- \partial_- \rho) \partial_- \partial_- \rho \partial_+ \rho \right)  \\
         &+ c \partial_- \left( F^\prime (\partial_+ \partial_+ \rho \ \partial_- \partial_- \rho) \partial_+ \partial_+ \rho \partial_- \rho\right)  \quad + \quad \mathcal{O}(c^2)\, . \nonumber 
\end{align}
As we will show in the next section, this equation is classically integrable for arbitrary choices of $F$.

\subsection{Lax integrability}
\subsubsection{The linear system}
\label{eq:GroupTheoreticArgument}
In order to prove the Lax integrability of \eqref{eq:Ldef} we employ the definition \cite{Ferko:2024ali} of the integrability of a Lagrangian deformed by auxiliary fields: the existence of a Lax connection whose flatness, \textit{upon imposition of the algebraic equations of motion}, is equivalent to the remaining Euler-Lagrange equations for the physical fields. This definition is interchangeable with the more general statement of complete equivalence between the flatness of a Lax connection and the full set of Euler-Lagrange equations (algebraic ones included), but it comes with the additional advantage of simplifying consistently many cumbersome expressions. In the following, we will refer to equalities that hold when equations \eqref{eq:algebraicP}, \eqref{eq:algebraicsigma} and \eqref{eq:algebraicrho} are satisfied via the symbol ``$\doteq$".

An inspection of the Lax connection $\mathfrak{L}_\mu(z)$ in \eqref{eq:largeLS} compared to the augmented ones for auxiliary field deformed models \cite{Ferko:2024ali}, \cite{Cesaro:2024ipq}, \cite{Bielli:2024oif}, leads us to posit the following natural ansatz for the Lax connection of the system \eqref{eq:Ldef}:
\begin{equation}\label{eq:Laxconnectiondef}
\tilde{\mathfrak{L}}_\mu=Q_\mu+\frac{1+z^2}{1-z^2}P_\mu+\frac{2z}{1-z^2}\epsilon_{\mu\nu}\mathcal{P}^\nu\,+\left(\frac{1+z^2}{1-z^2}\frac{\partial_\mu\rho}{\rho}+\frac{2z}{1-z^2}\epsilon_{\mu\beta}\mathcal{R}^\beta\right)z\frac{\partial}{\partial z}.
\end{equation}
The proof of the equivalence between
\be
2\partial_{[\mu}\tilde{\mathfrak{L}}_{\nu]}+[\tilde{\mathfrak{L}}_{\mu}, \tilde{\mathfrak{L}}_{\nu}]\doteq 0\, ,
\ee
and \eqref{eq:Boxrho=smth}, \eqref{eq:dynamicaleom},  is furnished in appendix \ref{appendix:weakintegrability}, and crucially relies on the properties
\begin{equation}
\epsilon^{\mu\nu}\partial_\mu\rho P_\nu \doteq \epsilon ^{\mu\nu}\mathcal{R}_{\, \mu}\mathcal{P}_\nu ,\quad \partial^\mu \rho \mathcal{P}_\mu \doteq \mathcal{R}^{\mu}P_\mu \, ,
\end{equation}
which can in turn be proven with the aid of \eqref{eq:epsK}. Nonetheless, this gives us no information about equation \eqref{eq:Boxsigma+smth}, which, compared to its undeformed counterpart \eqref{eq:boxsigma0}, is far from being trivially integrable; neither does it provide us with any insights about the nature of the new Virasoro constraints, ultimately encoding the diffeomorphism invariance of the theory.

\subsubsection{Virasoro constraints}
With the aim of addressing these conundrums, we reformulate the generalised linear system \eqref{eq:largeLS}
\begin{align}
\label{eq:largeLSdef}
 \hat{\mathcal{Y}}^{-1}\partial_\mu\hat{\mathcal{Y}}&= \left(\mathbf{\Gamma}^{-1}\hat{\mathcal{V}}^{-1}\partial_\mu\left(\hat{\mathcal{V}}\mathbf{\Gamma}\right), \, \left[\tilde{\mathcal{S}}_\mu-\Omega^{\prime}(\hat{\mathcal{V}}, \hat{\mathcal{V}}^{-1}\partial_\mu \hat{\mathcal{V}})\right]\mathsf{K}\right)\cong(\tilde{\mathfrak{L}}_{\mu}(z),0)\, ,
\end{align}
this time requiring that the imposition of the same twisted self-duality constraints \cite{Julia:1996nu} allows for the recovery of the deformed Lax connection \eqref{eq:Laxconnectiondef}, and where $\tilde{\mathcal{S}}_\mu$ is a new central element, different from $\partial_\mu\hat{\sigma}$. This is ultimately possible because, as the algebraic structure of \eqref{eq:Laxconnectiondef} also reveals, the underlying symmetry  \eqref{eq:largercoset} is preserved by the deformation \eqref{eq:Ldef}. For the same reason, $\hat{\mathcal{Y}}$ can be explicitly parameterised as in the undeformed case.

We show in appendix \ref{appendix:Virasoro} that the vanishing of the last entry in \eqref{eq:largeLSdef} furnishes the Virasoro constraints (in lightcone coordinates \eqref{eq:lightonecoords}) 
\begin{equation}\label{eq:deformedVirasoro}
  \tilde{\mathcal{S}}_\pm=-\frac{1}{2}\rho\,\text{Tr}\left(\frac{v_\pm v_\pm}{\chi_{2\,\pm}}+ \frac{K_\pm{}^\mp v_\mp v_\mp}{\chi_{2\,\mp}}\right)\, .
\end{equation}
In \eqref{eq:deformedVirasoro}, $\tilde{\mathcal{S}}_\pm$ is a modified version of $\mathcal{S}_\pm$ (namely the lightcone components of $\mathcal{S}_\mu$ defined in \eqref{eq:J}), satisfying $\partial^\mu\tilde{\mathcal{S}}_\mu=\partial^\mu\mathcal{S}_\mu$. Therefore,
\begin{equation}
    \tilde{\mathcal{S}}_\pm=-\partial_\pm\tilde{\hat{\sigma}}-2\chi_{1\, \pm},
\end{equation}
with $\tilde{\hat{\sigma}}$ a specifically rescaled version of $\sigma$, whose explicit form we do not need,  and such that, for $K_\pm{}^\mp=0$, 
\begin{equation}
\tilde{\hat{\sigma}}\Big\lvert_{K_\pm{}^\mp=0}=\sigma+ \frac{1}{2}\text{ln}(\partial_+\rho\partial_-\rho).
\end{equation}
Notice that in the smooth limit $K_\pm{}^\mp \rightarrow 0$ the undeformed Virasoro constraints \eqref{eq:Vir2} are recovered. 

One might now inquire about the nature of the field equation \eqref{eq:Boxsigma+smth}, namely whether it is an independent one or whether it results from the deformed Virasoro constraints \eqref{eq:deformedVirasoro}, upon imposition of \eqref{eq:Boxrho=smth}, \eqref{eq:dynamicaleom} and \eqref{eq:Bianchis}, akind to what happens for the undeformed model. In appendix \ref{appendix:BoxsigmafromVirasoro} we show that the second instance is indeed the case; equation \eqref{eq:Boxsigma+smth}, therefore is not really an independent one, and the generalised linear system \eqref{eq:largeLSdef}, able to provide a theoretical justification of \eqref{eq:deformedVirasoro} and whose compatibility is equivalent to equations \eqref{eq:Boxrho=smth}, \eqref{eq:dynamicaleom} and \eqref{eq:Bianchis}, is enough to establish the Lax integrability of the model \eqref{eq:Ldef}.
 \subsubsection{The centrally extended Lax connection}
Before switching gears to the Hamiltonian analysis, here we comment on the observation that the same Lax integrability of the deformed model can be proved by using an alternative, centrally extended, Lax connection as in subesction \ref{subsec:Lax}. The Lax pair can simply be constructed from the undeformed case \eqref{eq:UndefLax} by the heurestic substitutions
\begin{equation}
\star P \rightarrow \mathcal{P}, \qquad \star \mathrm{d}\rho \rightarrow \star \mathcal{R}, \qquad \star \mathrm{d}\sigma \rightarrow \star \mathcal{S}\, .
\end{equation}
The alternative Lax connection for the Auxiliary Field Deformation is
\begin{equation}\label{eq:alternativeLax}
    \tilde{\mathfrak{L}}_{\text{ext}\,\mu} = Q_\mu + \frac{1+z^2}{1-z^2} P_\mu + \frac{2z}{1-z^2} \epsilon_{\mu\nu} \mathcal{P}^\nu + \left( \frac{1+z^2}{1-z^2} \frac{\partial_\mu\rho}{\rho} + \frac{2z}{1-z^2} \frac{ \epsilon_{\mu\nu}\mathcal{R}^\nu}{\rho} \right) L_0 +\epsilon_{\mu\nu} \mathcal{S}^\nu\, \mathsf{K}\, , 
\end{equation}
where we added a central charge term $\mathsf{K}$ in the connection. In exactly the same way as in subsection \ref{subsec:Lax}, we have now 
\begin{equation}\label{eq:centralcomms}
    \left[ \frac{1+z^2}{1-z^2} P_\mu ,  \frac{2z}{1-z^2} \epsilon_{\nu\gamma}\mathcal{P}^\gamma \right] = [P_\mu,\epsilon_{\nu\gamma}\mathcal{P}^\gamma]\, \frac{2z (1+z^2)}{(1 - z^2)^2}+\frac{1}{4}\epsilon_{\nu\gamma} \text{Tr} \left( P_\mu \mathcal{P}^\gamma \right) \mathsf{K}\, .
\end{equation}
Flatness of \eqref{eq:alternativeLax} will be equivalent to the equations of motion \eqref{eq:dynamicaleom}, \eqref{eq:Boxrho=smth} and \eqref{eq:Boxsigma+smth}, together with the Bianchi identities \eqref{eq:Bianchis} (we adopt form notation in order to avoid cluttering once again),
\begin{align}
    0 &\doteq \mathrm{d}  \tilde{\mathfrak{L}}_{\text{ext}}+ \frac{1}{2} [  \tilde{\mathfrak{L}}_{\text{ext}} ,  \tilde{\mathfrak{L}}_{\text{ext}} ] \\
    &\doteq \left( \mathrm{d} Q + \frac{1}{2} [Q,Q] + \frac{1}{2} [P,P] \right) + \frac{1+z^2}{1 - z^2} \left(  \mathrm{d}P + [Q,P] \right) + \frac{2z}{1-z^2} \left( \mathrm{d} \star \mathcal{P} + [Q, \star \mathcal{P}] + \frac{\mathrm{d} \rho}{\rho} \wedge \star \mathcal{P} \right) \nonumber \\
    &{} \qquad + \frac{2z}{1-z^2} \left( \mathrm{d} \star \mathcal{R} \right) L_0 + \left( \mathrm{d} \star \mathcal{S} + \frac{1}{2} \text{Tr} (P \wedge \star \mathcal{P}) \right) \mathsf{K}\, , \nonumber
\end{align}
where we have used \eqref{eq:centralcomms} and the fact that there exists no invariant bilinear pairing between $L_0$ and $T_A$.

%
%
%
%

\subsection{Hamiltonian analysis}
The canonical momenta associated to $\{\mathcal{V}, \rho, \sigma\}$ are:
\begin{align}
    \pi^{(P)} &= \rho \mathcal{P}_t = - \rho ( P_t + 2 v_t), \\
    \pi^{(\rho)} &= \mathcal{S}_t = -  ( \partial_t \sigma + 2 \chi_{1t}), \\
    \pi^{(\sigma)} &= \mathcal{R}_t = -  ( \partial_t \rho + 2 \chi_{2t}).
\end{align}
In Hamiltonian formalism the Lax matrix becomes: 
\begin{equation}
    \mathfrak{L}_x = Q_x + \frac{1+z^2}{1-z^2} P_x + \frac{2z}{1-z^2} \frac{\pi^{(P)}}{\rho} + \left( \frac{1+z^2}{1-z^2} \frac{\partial_x\rho}{\rho} + \frac{2z}{1-z^2} \frac{ \pi^{(\sigma)}}{\rho} \right) L_0 + \pi^{(\rho)} \mathsf{K}.\label{eq:AFDefLaxHam}
\end{equation}
The Hamiltonian analysis of the action is complicated by the fact there are six constraints $\Phi^I = (\Phi^{(v)},\Phi^{(\chi_1)},\Phi^{(\chi_2)}) \approx 0$, associated to the algebraic equations of motion for $v,\chi_1, \chi_2$ and the canonical momenta for these, which vanish:
\begin{equation}
    \Psi^{(v)} \approx 0, \quad \Psi^{(\chi_1)} \approx 0, \quad \Psi^{(\chi_2)} \approx 0.
\end{equation}
The constraints $\Phi^I$ are explicitly given by:
\begin{align}
    \Phi_t^{(v)} &= \frac{1}{\rho} \pi^{(P)} - ((1 + K)v)_t, \qquad \Phi_x^{(v)} = P_x + ((1 - K)v)_x , \\
    \Phi_t^{(\chi_1)} &= \pi^{(\sigma)} - ((1 + K)\chi_2)_t, \qquad \Phi_x^{(\chi_1)} = \partial_x \rho + ((1 - K)\chi_2)_x , \\
    \Phi_t^{(\chi_2)} &= \pi^{(\rho)} - ((1 + K)\chi_1)_t, \quad \Phi_x^{(\chi_2)} = \partial_x \sigma + ((1 - K)\chi_1 )_x.
\end{align}
The constraints do not commute. Schematically, as in \cite{Cesaro:2024ipq}, the constraint matrix for the 6 commuting constraints $\Psi_I$, and the six generically non-commuting constraints $\Phi_I$ looks as follows
\begin{align}
    \left\lbrace \psi_i, \psi_j \right\rbrace = M_{ij} = \begin{pmatrix}
        0 & *\\
        * & *
    \end{pmatrix}\, ,
\end{align}
with $\psi_1=\Psi$ and $\psi_2=\Phi$. About the inverse, we need only the fact that
\begin{align}
(M^{-1})^{ij} =\begin{pmatrix}
    * & * \\
    * & 0
\end{pmatrix}    \,.
\end{align}
Remarkably, this means that the Dirac bracket for the physical fields corresponds to their Poisson brackets 
\begin{align*}
    &{} \quad \left \lbrace (P_x , \pi^{(P)}, \rho, \pi^{(\rho)}, \sigma , \pi^{(\sigma)}), (P_x , \pi^{(P)}, \rho, \pi^{(\rho)}, \sigma , \pi^{(\sigma)}) \right\rbrace_{D.B.}  \nonumber \\
    &= \left \lbrace (P_x , \pi^{(P)}, \rho, \pi^{(\rho)}, \sigma , \pi^{(\sigma)}), (P_x , \pi^{(P)}, \rho, \pi^{(\rho)}, \sigma , \pi^{(\sigma)}) \right\rbrace_{P.B.} \\
    &{} \quad - \left\lbrace (P_x , \pi^{(P)}, \rho, \pi^{(\rho)}, \sigma , \pi^{(\sigma)}), \Phi \right\rbrace_{P.B.} (M^{-1})^{\Phi\Phi} \left\lbrace \Phi, (P_x , \pi^{(P)}, \rho, \pi^{(\rho)}, \sigma , \pi^{(\sigma)}) \right\rbrace_{P.B.} \\
    &= \left \lbrace (P_x , \pi^{(P)}, \rho, \pi^{(\rho)}, \sigma , \pi^{(\sigma)}), (P_x , \pi^{(P)}, \rho, \pi^{(\rho)}, \sigma , \pi^{(\sigma)}) \right\rbrace_{P.B.}
\end{align*}
due to $(M^{-1})^{\Phi\Phi} = 0$. This implies that for the Lax matrix \eqref{eq:AFDefLaxHam}
\begin{align}
    \{ \mathfrak{L}_x(z,x) , \mathfrak{L}_x (z^\prime,x^\prime) \}_{D.B.} = \{ \mathfrak{L}_x(z,x) , \mathfrak{L}_x (z^\prime,x^\prime) \}_{P.B.}.
\end{align}
As in the other cases of such Auxiliary Field Deformations the canonical analysis of integrability, i.e. the canonical form of the Lax matrix and the involution of the charges, is not affected by the deformation and agrees with the expression in the undeformed case \eqref{eq:ExtLaxUndeformedCanonical}. The above argument means that the auxiliary field model has the same classical $r$-matrix in comparison to the undeformed model \eqref{eq:ExtLaxRmatrix}.
\section{Yang-Baxter deformation}
\label{sec:YB}

\subsection{Yang-Baxter deformation of $D=2$ dimensionally reduced gravity}

Integrable $\sigma$-models are quite sparse. For a long time, only $\sigma$-models with very particular and symmetric target spaces -- flat space, Lie (super)groups, symmetric spaces -- have been found to be integrable. A typical route to obtain new integrable models is to deform such known models at the level of the action. One example of non-trivial deformations has been introduced and investigated in the last decade via the so-called Yang-Baxter deformations \cite{Klimcik:2002zj,Klimcik:2008eq,Delduc:2013qra,Kawaguchi:2014qwa,Hollowood:2014rla,Kawaguchi:2014fca,Matsumoto:2015jja,Hoare:2016hwh}. These deformations are generated by so-called $R$-operators -- algebraic objects based on an underlying Lie algebra structure: solutions to the (modified) classical Yang-Baxter equation. In the former case these are typically called \textit{homogeneous Yang-Baxter deformation}, in the latter \textit{$\eta$-deformation}. Homogeneous Yang-Baxter deformations have been shown to be closely connected to abelian and non-abelian $T$-duality transformations \cite{Matsumoto:2014nra,vanTongeren:2015uha,Osten:2016dvf,Hoare:2016wsk,Borsato:2016pas,Borsato:2017qsx}. $\eta$-deformations, on the other hand, are non-trivial deformations of the classical dynamics and have novel target space interpretations \cite{Delduc:2013fga,Delduc:2014kha,Hoare:2015wia,Arutyunov:2015mqj,Hoare:2016ibq,Borsato:2016ose,Orlando:2016qqu,Borsato:2018spz,Hoare:2018ngg,vanTongeren:2019dlq,Osten:2021opf} in the case of $\sigma$-models. By now, Yang-Baxter deformations have also been shown to fit into the landscape of classical Gaudin models and 4d Chern-Simons theory \cite{Delduc:2019whp,Lacroix:2023gig,Levine:2023wvt}. 

So far, Yang-Baxter deformations have been considered mainly applied to string $\sigma$-models. Here, we introduce the following Lagrangian as the Yang-Baxter deformation of $D=2$ dilaton-gravity coupled to the coset $\sigma$-model\footnote{Again, for the sake of integrability, we can choose a generic symmetric space $G/K$ as target space. Concrete examples of dimensionally reduced gravity correspond to specific choices of $G/K$, as mentioned in section \ref{sec:Review}.}, \eqref{eq:LagEgconf}: 
\begin{equation}
    L = \frac{1}{\beta(\rho)} (\partial_+ \rho \partial_- \sigma + \partial_- \rho \partial_+ \sigma) - \frac{\rho}{2} \text{tr} \left( P_+ \frac{1}{1 - \rho R_g \circ \mathbb{P}_\mathfrak{p}} P_- \right) \, .\label{eq:YB2dGrav}
\end{equation}
In \eqref{eq:YB2dGrav}, $\mathbb{P}_\mathfrak{p}$ is the projector onto the $\mathfrak{p}$ part of $\mathfrak{g}$, $Q + P = \mathcal{V}^{-1} \mathrm{d} \mathcal{V}$ and $R_g = \text{Ad}_{g}^{-1} \circ R \circ \text{Ad}_g$ with the operator $R: \ \mathfrak{g} \rightarrow \mathfrak{g}$ being a solution of the (modified) classical Yang-Baxter equation
\begin{equation}
    [R(m),R(n)] - R([R(m),n] - [m,R(n)]) = - c^2 C^2 [m,n] ,\label{eq:MCYBE}
\end{equation}
for $m,n\in \mathfrak{g}$. The choices $c^2 = 0,1,-1$ correspond to homogeneous, split and non-split case, see \cite{Vicedo:2015pna} for a review. In particular, the space of solutions of this equation and, hence, also of possible Yang-Baxter deformations is unchanged in comparison to pure $\sigma$-model case. 

The constant $C$ in \eqref{eq:MCYBE} comes from the identification $\eta = C\rho$, to be justified below, in comparison to the standard form of the Yang-Baxter deformation (for the $\sigma$-model part):
\begin{equation}
    L = - \frac{\rho}{2} \text{tr} \left( P_+ \frac{1}{1 - \eta \tilde{R}_g \circ \mathbb{P}_{\mathfrak{p}}} P_- \right)\, ,
\end{equation}
with $ \tilde{R} = C R$. Let us explain why we assume $\eta = C \rho$. This condition is necessary s.t. the Maurer-Cartan identity in terms of the deformed current takes the form \eqref{eq:MaurerCartanIdentityDeformed1} similar to the standard Yang Baxter $\sigma$-model, see \eqref{eq:YBPCMMaurerCartan} in appendix \ref{appendix:PCM2dgravity}  for the corresponding calculation for the Principal Chiral Model. This requires $\mathrm{d}\eta = \eta \frac{\mathrm{d} \rho}{\rho}$ which has three independent solutions, of which the last one will be the only interesting one here:
\begin{enumerate}
    \item $\eta = 0$. This corresponds to the standard undeformed dimensionally reduced gravity.
    \item $\mathrm{d} \eta = \mathrm{d} \rho = 0$. This would correspond to the standard Yang-Baxter deformation of the coset $\sigma$-model.
    \item $\eta = C \rho$ for some real constant $C$. This factor of $C$ is absorbed into the $R$-operator defining the deformation, and hence appears in the present version of the modified classical Yang-Baxter equation \eqref{eq:MCYBE}. It is this last case, that we are interested in. As a consequence, the non-deformation limit $C \rightarrow 0$ at the level of the Lagrangian is given by $R_g \rightarrow 0$. 
\end{enumerate}
As we will see, the function $\beta(\rho)$ will be fixed to be
\begin{equation}
    \beta(\rho) = 1 - c^2 C^2 \rho^2 
\end{equation}
for the model to be integrable. The integrability of this particular model (Yang-Baxter deformation of the coset $\sigma$-model with spacetime dependent couplings) had not been introduced in \cite{Hoare:2020fye}. Moreover, here we extended their setting to the coupling to full-fledged $D=2$ gravity, giving a physical meaning to the spacetime dependent coupling constant as the dilaton.

In appendix \ref{appendix:PCM2dgravity}, we introduce the corresponding calculation for the Yang-Baxter deformation of the Principal Chiral Model coupled to $D=2$ gravity, suggesting that this procedure is feasible for all integrable models with `running couplings' introduced in \cite{Hoare:2020fye}.

\subsection{Equations of motion}
The equations of motion to the action \eqref{eq:YB2dGrav} are
\begin{align}
    \mathrm{d} \left( \frac{\star \mathrm{d} \rho }{\beta(\rho)} \right) &= 0 \, ,\label{eq:YBEquationOfMotionRho} \\
    \mathrm{d} \star \mathrm{d} \sigma &= - \frac{\beta(\rho)}{2} \mathrm{Tr}( K^{(1)} \wedge \star K^{(1)} )\;,  \label{eq:YBEquationOfMotionSigma} \\
    \mathrm{D} \star (\rho K^{(1)}) &= 0  \quad \rightarrow \quad \mathrm{D} \star K^{(1)} = - \frac{\mathrm{d} \rho}{\rho} \wedge K^{(1)}\, , \label{eq:YBEquationOfMotionK}
\end{align}
in terms of the deformed current $K$, which is defined by:
\begin{equation}
    J = Q + P = K + \eta \star R_g K^{(1)}.
\end{equation}
This ensures that:
\begin{equation}
   K_\pm = \frac{1}{1 \pm \rho R_g \circ \mathbb{P}_{\mathfrak{p}}} P_\pm.
\end{equation}
Remarkably, the second equation \eqref{eq:YBEquationOfMotionSigma} takes this form from the variation with respect to $\rho$, also in its new role as a deformation parameter in \eqref{eq:YB2dGrav}. This equation is a novel feature in comparison to the cases of the other integrable models in \cite{Hoare:2020fye}, where the `running coupling' $\rho$ was not considered to be a dynamical field.

The Maurer-Cartan equation of $J$ in terms of $K$ takes the same form as in the standard Yang-Baxter deformed $\sigma$-model case:
\begin{align}
    0 &= \mathrm{d} Q + \frac{1}{2} [Q,Q] + \frac{1}{2} [P,P] = \mathrm{d} K^{(0)} + \frac{1}{2} [K^{(0)},K^{(0)}] + \frac{1}{2} (1 - c^2 C^2 \rho^2) [K^{(1)},K^{(1)}], \label{eq:MaurerCartanIdentityDeformed1} \\
    0 &= \mathrm{d} P + \frac{1}{2} [Q,P]  = \mathrm{d} K^{(1)} + [K^{(0)},K^{(1)}] .\label{eq:MaurerCartanIdentityDeformed2}
\end{align}
These are valid on-shell, i.e. in particular if \eqref{eq:YBEquationOfMotionK} holds.

\subsection{Lax connection}
As in \cite{Hoare:2020fye}, we make the same ansatz for the Lax pair as in the case with constant couplings\footnote{Deformation parameter and constant in front of $\sigma$-model action.}, namely
\begin{align}
    \mathfrak{L}(\lambda)_{Y.B.} = K^{(0)} + \sqrt{\beta(\rho)} \left( \frac{1+\gamma^2}{1 - \gamma^2} K^{(1)} + \frac{2\gamma}{1 - \gamma^2} \star K^{(1)} \right) .\label{eq:YBLaxConnectionMatter}
\end{align}
This ansatz only captures the dynamics of the matter sector, but typically is expected to reproduce the equation of motion for $\rho$ \eqref{eq:YBEquationOfMotionRho} as integrability condition. We will see that the same is true here.

This Lax connection is flat, given that $K$ satisfies the equation of motion, Maurer-Cartan identity and the spectral parameter $\gamma$ is dynamical and subject to the following condition:
\begin{equation}
    A(\gamma) B(\gamma) \frac{\mathrm{d} \gamma}{\gamma} - B^2(\gamma) \frac{\star \mathrm{d} \gamma}{\gamma} = \frac{B(\gamma)}{\beta(\rho)} \frac{\mathrm{d} \rho}{\rho} - A(\gamma) \frac{c^2 C^2 \rho^2}{\beta(\rho)} \frac{\star \mathrm{d} \rho}{\rho}\,,
\end{equation}
with
\begin{equation}
    A(\gamma) = \frac{1+\gamma^2}{1 - \gamma^2}, \quad B(\gamma) = \frac{2 \gamma}{1 - \gamma^2}, \quad \beta(\rho) = 1 - c^2 C^2 \rho^2 \,,
\end{equation}
or equivalently:
\begin{equation}
    \frac{\mathrm{d} \gamma}{\gamma} = A(\gamma) \frac{\mathrm{d}\rho}{\rho} + \left( B(\gamma) - \frac{c^2 C^2 \rho^2}{B(\gamma) \beta(\rho)} \right) \frac{\star \mathrm{d}\rho}{\rho}.
\end{equation}
Integrability of this equation requires
\begin{equation}
    \mathrm{d} \star \mathrm{d} \rho + \frac{2c^2 C^2 \rho}{1 - c^2 C^2 \rho^2} \mathrm{d} \rho \wedge \star \mathrm{d}\rho = 0.
\end{equation}
This reproduces the form (3.10) of \cite{Hoare:2020fye}
\begin{equation}
    \mathrm{d} \star \mathrm{d} \rho - \frac{\beta^\prime(\rho)}{\beta(\rho)} \mathrm{d} \rho \wedge \star \mathrm{d} \rho = 0, \label{eq:DilatonYB}
\end{equation}
for $\beta(\rho) = 1 - c^2 C^2 \rho^2$, which is in fact the $\beta$-function for the deformation parameter $\eta$ (here $\rho$) \cite{Hoare:2019ark}. This can be equivalently written as
\begin{equation}
    \mathrm{d} \star \left( \frac{\mathrm{d} \rho}{\beta(\rho)} \right) = 0.
\end{equation}
This integrability condition coincides with the equation of motion \eqref{eq:YBEquationOfMotionRho} for $\rho$.

Following the arguments in \cite{Hoare:2020fye}, generalising the argument in the standard $D=2$ undeformed model, one can choose coordinates $(t,x)$ such that $\rho = \rho(t)$ and
\begin{equation}
    \partial_t \rho = \rho(t).
\end{equation}
This implies that in suitable coordinates $\rho$ can have non-trivial dynamics along the RG-flow with world-sheet time being RG-time.

A Lax connection with constant spectral parameter, analogous to \eqref{eq:alternativeLax} does \textit{not} seem to exist. The problem is that the following natural choice 
\begin{equation}
    \mathfrak{L} = K^{(0)} + \sqrt{\beta(\rho)} \left( A(z) K^{(1)} + B(z) \star K^{(1)} \right) + \left( A(z) \frac{\mathrm{d} \rho}{\rho} + B(z) \star \frac{1}{\beta(\rho)} \frac{\mathrm{d} \rho}{\rho} \right) L_0 + \star \mathrm{d}\sigma \mathsf{K} \nonumber
\end{equation}
reproduces the correct $\rho$ and $\sigma$ equations of motion, but \textit{not} the one of $K$.

The reason for this could be that, contrary to the Auxiliary Field Deformation, the Yang-Baxter one deforms the underlying group theoretic structure of the model in a more significant way. In particular, it is known that the underlying structure of the Yang-Baxter deformation is the one of a Lie bialgebra\footnote{For reductions of $D=4$ General Relativity, $\mathfrak{g}$ would be $\mathfrak{sl}(2)$ here.} $\mathfrak{d} = \mathfrak{g} \oplus \mathfrak{g}^\star$ associated with the solution $R$ to the modified classical Yang-Baxter deformation, that is used to define the deformation \cite{Vicedo:2015pna}. The group theoretic arguments in \ref{eq:GroupTheoreticArgument} in the auxiliary field case relied on the fact that the same group theoretic structure holds as in the undeformed case \ref{sec:Laxintegrability}.

For the same reason, we also do not comment on the Hamiltonian integrability of the Yang-Baxter deformed $D=2$ gravity model here. Whereas in the auxiliary field case the canonical structure of most of the variables does not change in comparison to the undeformed case, for the Yang-Baxter deformation, even in the pure $\sigma$-model case, the classical $r$/$s$-matrices change, see \cite{Lacroix:2018njs} for a review on this subject. In case of $D=2$ dimensionally reduced gravity, this would necessitate a totally new analysis of the Poisson involution as performed in \cite{Samtleben:1998fk} and goes beyond the scope of the present paper and is left for future work. As commented on already in \cite{Cesaro:2024ipq}, the fact that the auxiliary deformation does not change the classical $r$/$s$-matrix indicates that it is a rather `mild' deformation (as are the $T\bar{T}$-deformations), in comparison to the (inhomogeneous) Yang-Baxter deformations.

\section{Outlook}
\label{sec:Outlook}
In this paper, we have provided new integrable deformations of the dimensional reduction of $D=4$ General Relativity along two commuting Killing isometries. For the Auxiliary Field Deformation \eqref{eq:Ldef}, we have proven both Lax and Hamiltonian integrability, in part making use of the rich symmetry structures threading the original $D=2$ theory, which, remarkably, are preserved by the deformed one as well. For the Yang-Baxter deformation, we have provided a flat Lax representation which proves its Lax integrability; therefore, successfully incorporating and extending the analysis of \cite{Hoare:2020fye} to the dimensional reduction of $D=4$ General Relativity. 

Our results do not only supply new integrable models by themselves, but they feature the interplay of gravitational interaction with dilaton and coset-like matter, and therefore represent new arenas where to test (exact) solution generating techniques. Nonetheless, the same results also lead to interesting further questions, which remain so far open and which we here wish to comment on.

To begin with, it is surely tempting to speculate about the possible higher dimensional origin of both models \eqref{eq:Ldef}, \eqref{eq:YB2dGrav}. For what concerns \eqref{eq:Ldef}, the  Lagrangian seems unlikely to solely arise from a (matter) energy-momentum deformation of the theory in $D=3$ (it would not be immediate to incorporate derivatives of $\sigma$); but we could inquire whether it can instead be traced back to the presence of higher order derivative terms deforming the theory either in $D=3$ or in $D=4$. 
This would be surprising, at least for \eqref{eq:Ldef}:  higher order derivatives in higher dimensional theories of gravity are assumed to break in general the hidden symmetry group of the dimensionally reduced model~\cite{Michel:2007vh, Lambert:2006he, Lambert:2006ny, Bao:2007er} (with the exception of trivial field redefinitions, see \cite{Colonnello:2007qy}), whereas we have ascertained the presence not only of the full Geroch group, but even of the enlarged symmetry \eqref{eq:largercoset}. 
An intriguing  way out could be the presence of only a certain, possibly infinite dimensional subset of the infinite tower of higher derivative terms correcting the theory in $D=4$ or $D=3$, conspiring to preserve the lower dimensional duality symmetry. This is certainly a possibility that deserves attention, and we here leave it open to future investigations. 

The physical interpretation of the Yang-Baxter deformed dilaton-gravity models in section \ref{sec:YB} is also left as an open question for future work. We expect that a higher dimensional origin on \eqref{eq:YB2dGrav} in terms of $D=4$ Chern-Simons theory will exist, as both the Yang-Baxter deformations of $\sigma$-models and undeformed $D=2$ gravity fit into this framework \cite{Delduc:2019whp,Cole:2024skp}. A higher-dimensional origin of the $D=2$ model as some reduction of a higher dimensional gravity theory seems highly speculative. In particular, the Yang-Baxter deformation would generically lead to $B$-field terms (in the language of string $\sigma$-models), that typically do not appear in $D=2$ dimensionally reduced (super)gravity theories. 

We expect that all integrable deformations of symmetric space $\sigma$-models -- bi-Yang-Baxter \cite{Delduc:2014uaa,Klimcik:2014bta,Hoare:2014oua,Delduc:2015xdm}, Wess-Zumino terms, $\lambda$-deformations \cite{Sfetsos:2013wia} and combinations of those \cite{Klimcik:2017ken,Delduc:2017fib,Delduc:2018xug,Klimcik:2019kkf,Seibold:2019dvf,Hoare:2020mpv} -- could also be embedded in $D=2$ dimensionally reduced gravity theories, generalising the results in \cite{Hoare:2020fye}, as for all of these a modified integrability condition for $\rho$ of the form \eqref{eq:DilatonYB} has been found for different choices of $\beta(\rho)$. Equations of motion of the form \eqref{eq:DilatonYB} can always be obtained from the $D=2$ gravity part of the Lagrangian like $\frac{\mathrm{d}\rho}{\beta(\rho)} \wedge \star \mathrm{d}\sigma$. A common description of all of these models in the context of $D=4$ Chern-Simons theory, as in \cite{Cole:2024skp} for the undeformed $D=2$ dimensionally reduced gravity, would be interesting, as well.

A key question, also partly motivating this work, was whether it is possible to extend the techniques of integrable $D=2$ field theories to cases with dilaton potentials, i.e.
\begin{equation}
    \Box \rho = - V(\rho),
\end{equation}
with a scalar potential $V$. In this context, (pure) AdS$_2$ dilaton-gravity with non-trivial potentials  giving rise to solvable models, and even Yang-Baxter deformations thereof, have been studied \cite{Almheiri:2014cka,Kyono:2017jtc,Kyono:2017pxs
}; we stress however that these are simpler examples without coset matter coupling to $D=2$ gravity, as the ones analysed in this paper. 

Neither of the deformations considered in this article did lead to such a scalar potential: the Auxiliary Field Deformation sources \textit{higher derivative} modifications \eqref{eq:DilatonAUX} and the Yang-Baxter deformations lead to a modified equation for the dilaton, which is \textit{second order in derivatives} \eqref{eq:DilatonYB}. It would therefore be compelling to understand if other kinds of integrable deformations could lead to such scalar potentials instead. This would provide new opportunities to study cosmologies in $D=2$ with the methods of integrability.

\subsubsection*{Acknowledgements}

We would like to thank Olaf Hohm, Axel Kleinschmidt, Benedikt K\"onig, Sucheta Majumdar and Hermann Nicolai for useful discussions. We are particularly indebted to Axel Kleinschmidt for collaboration during the initial stage of this project and for carefully reviewing the manuscript. MC acknowledges the warm hospitality of the  Institute of Theoretical Physics at the University of Wrocław, where part of the project was carried out. The research of DO is part of the project No. 2022/45/P/ST2/03995 co-funded by the National Science Centre and the European Union’s Horizon 2020 research and innovation programme under the Marie Sk\l odowska-Curie grant agreement no.~945339.

\vspace{10pt}
\includegraphics[width = 0.09 \textwidth]{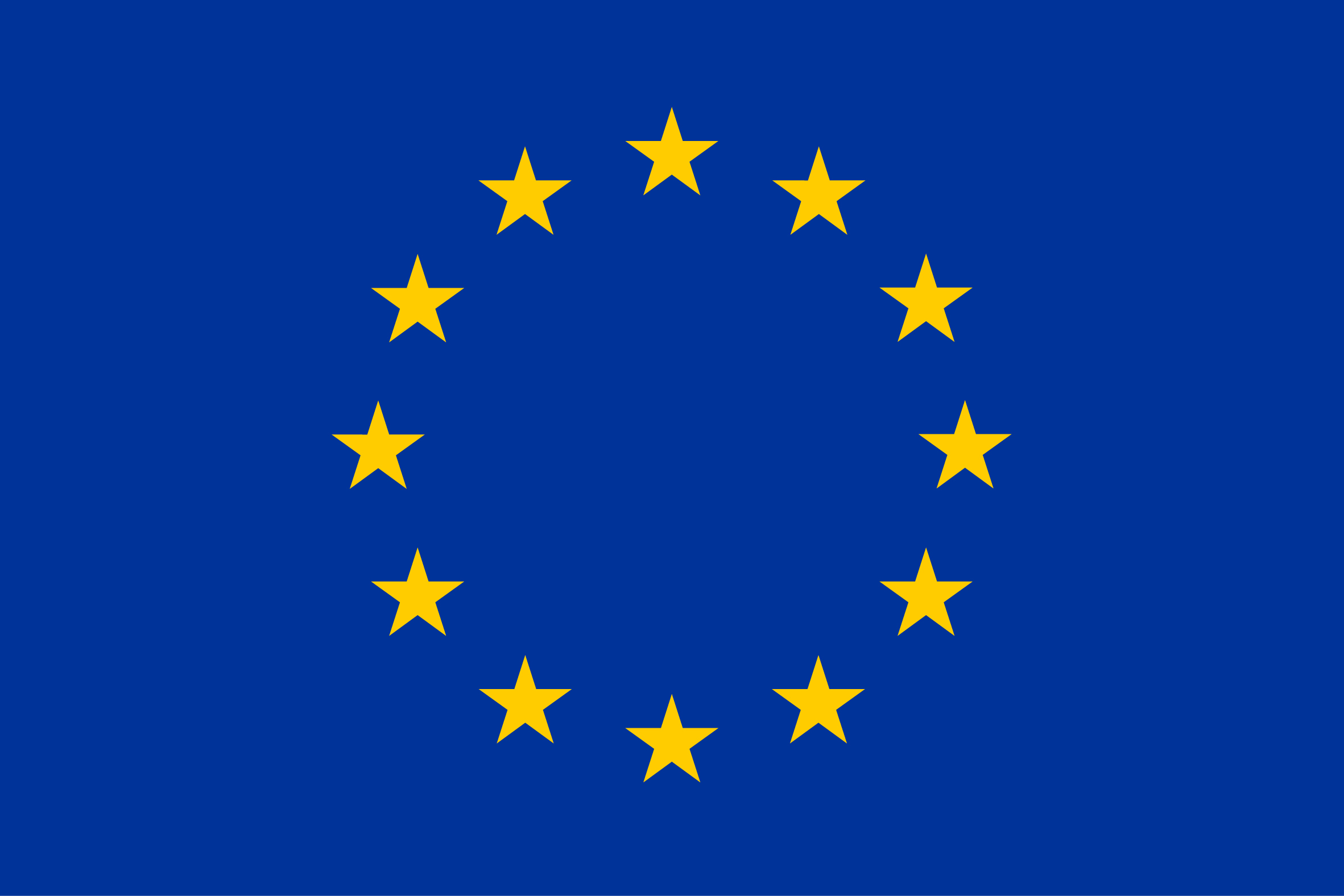} $\quad$
\includegraphics[width = 0.7 \textwidth]{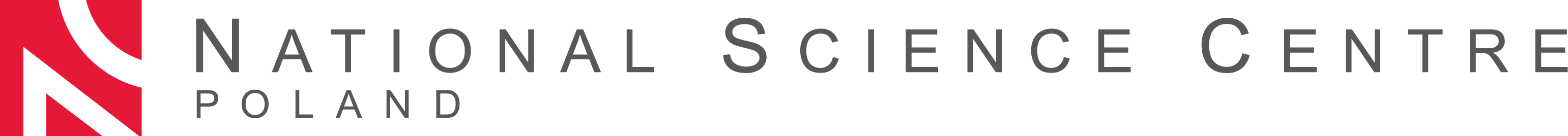}

\appendix
\section{Details on $\widehat{\text{SL}(2)}$}
\label{app:cosetmodels}

We here provide some details on $\widehat{\text{SL(2)}}$, and to its extension to $\mathcal{W} \ltimes \widehat{\text{SL(2)}}$ described in section \ref{sec:Laxintegrability}. Let us denote the $G=\text{SL(2)}_{(\text{E})}$ generators as $T^A$, with $A=1,2,3$, satisfying
\be
[T^A, T^B]=f^{AB}{}_CT^C\, ,
\ee
and where $f^{AB}{}_{C}$ are the SL(2) structure constants. The loop generators of $\widehat{\text{SL(2)}}$ are then denoted by $T_m^A$, with mode number $m\in\mathbb{Z}$, and obey the commutation relations
\be
[T_m^A, T_n^B]=f^{AB}{}_C T_{m+n}^C+m\eta^{AB}\delta_{m+n, 0}\mathsf{K}\, ,
\ee
with $\eta^{AB}$ the  Killing-Cartan 2-form and where $\mathsf{K}$ is the central extension of the loop algebra,
\be
[\mathsf{K}, T_m^A]=0\quad \forall \; m.
\ee
To complete $\widehat{\text{SL(2)}}$ we also need to define the derivation operator $\mathsf{d}$, satisfying
\be
[\mathsf{K},\mathsf{d}]=0\, ,\quad [\mathsf{d}, T_m^A]=-mT_m^A\, .
\ee
The Witt-Virasoro operators generating $\mathcal{W}$ 
obey the standard algebra
\begin{align}\label{eq:Virgens}
[L_m, L_n]&=(m-n)L_{m+n}\, ,\\
[L_m, T_n^{A}]&=-nT_{m+n}^A\, ,
\end{align}
from which we ascertain that $L_0\sim\mathsf{d}$, up to central terms. In terms of a constant spectral parameter $z$, all generators can be realised as
\be
T_m^A=z^mT^A,\quad L_m=-z^{m+1}\frac{\partial}{\partial z}\, .
\ee

\section{Lax integrability of the Auxiliary Field Deformation}\label{appendix:weakintegrability}
In order to check the flatness of \eqref{eq:Laxconnectiondef} we need
\begin{align}\label{eq:dLundef}
 2\epsilon^{\mu\nu}\partial_{\mu}\tilde{\mathfrak{L}}_{\nu}=&   2\epsilon^{\mu\nu}\partial_{\mu}Q_{\nu} +\frac{1+z^2}{1-z^2} 2\epsilon^{\mu\nu}\partial_{\mu}P_{\nu}+\frac{4z}{1-z^2}\epsilon^{\mu\nu}\partial_\mu\mathcal{P}_\nu+\frac{4z}{1-z^2}\left(\frac{\partial_\mu\mathcal{R}_\mu}{\rho}-\frac{\partial_\mu\rho\mathcal{R}^\mu}{\rho^2}\right)z\partial_z\, , 
\end{align}
and
\begin{align}\label{eq:LLundef}
   \nonumber \epsilon^{\mu\nu}[\tilde{\mathfrak{L}}_\mu, \tilde{\mathfrak{L}}_\nu]=&\epsilon^{\mu\nu}[Q_\mu, Q_\nu]+2\frac{1+z^2}{1-z^2}\epsilon^{\mu\nu}[Q_\mu, P_\nu]+\frac{4z}{1-z^2}[Q_\mu, \mathcal{P}^\mu]+\frac{(1+z^2)^2}{(1-z^2)^2}\epsilon^{\mu\nu}[P_\mu,P_\nu]\\[4pt]
   \nonumber & +\frac{4z(1+z^2)}{(1-z^2)^2}[P_\mu, \mathcal{P}^\mu]-\frac{4z^2}{(1-z^2)^2}\epsilon^{\mu\nu}[\mathcal{P}_\mu, \mathcal{P}_\nu]+2\epsilon^{\mu\nu}P_\mu\frac{\partial_\nu\rho}{\rho}\left[\frac{1+z^2}{1-z^2},\frac{1+z^2}{1-z^2}z\partial_z\right]\\[4pt]
   \nonumber &+2P_\mu\frac{\mathcal{R}^\mu}{\rho}\left[\frac{1+z^2}{1-z^2}, \frac{2z}{1-z^2}z\partial_z\right]-2\mathcal{P}_\mu\frac{\partial^\mu\rho}{\rho}\left[\frac{2z}{1-z^2}, \frac{1+z^2}{1-z^2}z \partial_z\right]\\
   &-2\epsilon^{\mu\nu}\mathcal{P}_\mu\mathcal{R}_\nu\left[\frac{2z}{1-z^2}, \frac{2z}{1-z^2}z\partial_z\right]+2\frac{\partial_\mu\rho\mathcal{R}^\mu}{\rho^2}\left[\frac{1+z^2}{1-z^2}z\partial_z, \frac{2z}{1-z^2}z\partial_z\right]\, .
\end{align}
After some algebraic manipulations we obtain
\begin{align}\label{eq:finalflatness}
\nonumber&\epsilon^{\mu\nu}\left(2\partial_{\mu}\tilde{\mathfrak{L}}_{\nu}+[\tilde{\mathfrak{L}}_\mu, \tilde{\mathfrak{L}}_\nu]\right)   = \\[4pt] 
 \nonumber&\epsilon^{\mu\nu}(2\partial_{\mu}Q_{\nu}+[Q_\mu, Q_\nu]+[P_\mu, P_\nu])+\epsilon^{\mu\nu}\frac{1+z^2}{1-z^2}(2\partial_{\mu}P_{\nu]}+[Q_\mu, P_\nu]+[P_\mu, Q_\nu])\\[4pt]
 \nonumber&+\frac{4z(1+z^2)}{(1-z^2)^2}[P_\mu, 2 v^\mu]+\frac{8z^2}{1-z^2}\epsilon^{\mu\nu}([v_\mu, P_\nu]+[P_\mu, v_\nu]+2[v_\mu, v_\nu])\\[4pt]
 \nonumber&+\frac{4z}{1-z^2}\left(\partial_\mu\mathcal{P}^\mu+[Q_\mu, \mathcal{P}^\mu]\right)+2\epsilon^{\mu\nu}\left(\mathcal{P}_\mu\mathcal{R}_\nu-P_\mu\frac{\partial_\nu\rho}{\rho}\right)\frac{4z^2(1+z^2)}{(1-z^2)^3}\\
 &+2\mathcal{P}_\mu\frac{\partial^\mu\rho}{\rho}\frac{4(1+z^2)^2z}{(1-z^2)^3}-P_\mu\frac{\mathcal{R}^\mu}{\rho}\frac{16z^3}{(1-z^2)^3}+\frac{4z}{1-z^2}\frac{\partial_\mu\mathcal{R}^\mu}{\rho}z\partial_z\, .
\end{align}
Flatness of \eqref{eq:finalflatness} should be equivalent to the equations of motion \eqref{eq:dynamicaleom} and \eqref{eq:Boxrho=smth}, together with the algebraic ones \eqref{eq:algebraicP}, \eqref{eq:algebraicrho}  and the Bianchi identities \eqref{eq:Bianchis}. The first line is indeed the two Bianchi identities for $Q_\mu$ and $P_\nu$. The second line of \eqref{eq:finalflatness} is zero iff 
\be
P_\mu=A_\mu{}^\nu v_\nu\, ,
\ee
where $A_\mu{}^\nu$ is a generic tensor, symmetric in its indices $(\mu,\nu)$ when both raised or lowered by the flat metric, endowed with the additional property
\be
\epsilon^{\mu\alpha}A_\alpha{}^\nu=-\epsilon^{\nu\alpha}A_{\alpha}{}^\mu.
\ee
For 
\be\label{eq:choiceofA}
A_\mu{}^\nu=-\delta_\mu^\nu+K_\mu{}^\nu\,,
\ee
this amounts indeed to the equation for the auxiliary fields \eqref{eq:algebraicP}, but, intriguingly, as already noticed in \cite{Cesaro:2024ipq}, it seems to allow for the possibility of much more general solutions.

The last element of the third line similarly vanishes iff
\be\label{eq:mixedchoice}
P_\mu=A_\mu{}^\nu v_\nu\, ,\quad \partial_\mu\rho=A_\mu{}^\nu \chi_{2\,\nu}
\, ,\ee
which for \eqref{eq:choiceofA} returns both and \eqref{eq:algebraicP} and \eqref{eq:algebraicrho}. The very same condition \eqref{eq:mixedchoice} leads to a recombination of the first two terms in the last line of \eqref{eq:finalflatness}, in such a way that
\begin{equation}
    \epsilon^{\mu\nu}\left(2\partial_{\mu}\tilde{\mathfrak{L}}_{\nu}+[\tilde{\mathfrak{L}}_\mu, \tilde{\mathfrak{L}}_\nu]\right)\doteq\frac{4z}{1-z^2}\left(D_\mu\mathcal{P}^\mu+\mathcal{P}^\mu\frac{\partial_\mu\rho}{\rho}\right)+\frac{4z}{1-z^2}\frac{\partial_\mu\mathcal{R}^\mu}{\rho} z\partial_z\, ,
\end{equation}
whose flatness is equivalent to equations \eqref{eq:dynamicaleom} and \eqref{eq:Boxrho=smth}.
\section{Virasoro constraint for the Auxiliary Field Deformation}
\subsection{Virasoro constraint from $\tau^\infty$ invariance of the linear system}\label{appendix:Virasoro}
We here follow the reasoning from \cite{Paulot:2004hh} in order to derive the Virasoro constraints from the vanishing of the $\mathsf{K}$-coefficient  in \eqref{eq:largeLSdef}. We reformulate such condition in terms of lightcone coordinates \eqref{eq:lightonecoords}
\be\label{eq:centralalg}
\tilde{\mathcal{S}}=\Omega^\prime\left(\hat{\mathcal{V}}^{-1}, Q+\left(\frac{1+\gamma^2}{1-\gamma^2}P_+-\frac{2\gamma}{1-\gamma^2}\mathcal{P}_+\right)dx^++\left(\frac{1+\gamma^2}{1-\gamma^2}P_-+\frac{2\gamma}{1-\gamma^2}\mathcal{P}_-\right)dx^-\right)\, ,
\ee
for some central factor $\tilde{\mathcal{S}}=\tilde{\mathcal{S}}_+dx^++\tilde{\mathcal{S}}_-dx^-$. In the proximity of some spacetime point $x_0$, we can locally write $\mathcal{V}=\mathbf{1}+\delta\hat{\mathcal{V}}$, and therefore re-express \eqref{eq:centralalg} as
\be
\tilde{\mathcal{S}}_\pm=\frac{1}{2}\omega\left(-\delta\hat{\mathcal{V}},\frac{1+\gamma^2}{1-\gamma^2}P_\pm \mp \frac{2\gamma}{1-\gamma^2}\mathcal{P}_\pm\right)\, ,
\ee
where $\omega$ is the algebra co-cycle. The latter is defined for two elements of $\text{Lie}(\widehat{\text{SL(2)}})$, $g_1(\gamma)$ and $g_2(\gamma)$, as follows
\be
\label{eq:algebracocyle}
\omega(g_1(\gamma), g_2(\gamma))=\frac{1}{2}\oint_{C_1}d\gamma\text{Tr}(\partial_\gamma g_1(\gamma) g_2(\gamma))+\frac{1}{2}\oint_{C_2}d\gamma\text{Tr}(\partial_\gamma g_1(\gamma) g_2(\gamma))
\ee
with $C_1$ and $C_2$ two contours exchanged and reversed under $\gamma \rightarrow \frac{1}{\gamma}$, avoiding singularities. According to this definition, we must take half of the residues around the simple poles $\gamma=\pm1$ of
\be
-\partial_\gamma\delta\hat{\mathcal{V}}\left(\frac{1+\gamma^2}{1-\gamma^2}P_\pm \mp \frac{2\gamma}{1-\gamma^2}\mathcal{P}_\pm\right)\, ,
\ee
returning ($\delta\hat{\mathcal{V}}$ is a regular function of $\gamma$)
\be\label{eq:Residuesintegral}
\tilde{\mathcal{S}}_\pm=\frac{1}{4}\text{Tr}\left[\mp\partial_\gamma\delta\hat{\mathcal{V}}\Big\lvert_{\gamma=-1}(P_\pm\pm\mathcal{P}_\pm)\mp\partial_\gamma\delta\hat{\mathcal{V}}\Big\lvert_{\gamma=+1}(P_\pm\mp\mathcal{P}_\pm)\right].
\ee
In order to handle the $\gamma$ derivatives of $\delta \hat{\mathcal{V}}$, we build on the observation \cite{Paulot:2004hh} that the poles in 
\be
Q+\frac{1+\gamma^2}{1-\gamma^2}\mathcal{P}+\frac{2\gamma}{1-\gamma^2}\star \mathcal{P}
\ee
only come from $\mathbf{\Gamma}(\mathbf{\Gamma}^{-1}d\mathbf{\Gamma})\mathbf{\Gamma}^{-1}\,\hat{\mathcal{V}}^{-1}\partial_\gamma\hat{\mathcal{V}}$, with
\be
d\mathbf{\Gamma}\mathbf{\Gamma}^{-1}=\left(\frac{1+\gamma^2}{1-\gamma^2}\frac{d\rho}{\rho}+\frac{2\gamma}{1-\gamma^2}\frac{\star d\mathcal{R}}{\rho}\right)\gamma\partial_\gamma\, .
\ee
Therefore, 
\begin{equation}
   \text{Res}\Big\lvert_{\pm 1}\left[\left(\frac{1+\gamma^2}{1-\gamma^2}\frac{d\rho}{\rho}+\frac{2\gamma}{1-\gamma^2}\frac{\star d\mathcal{R}}{\rho}\right)\gamma\hat{\mathcal{V}}^{-1}\partial_\gamma\hat{\mathcal{V}}\right] = \text{Res}\Big\lvert_{\pm 1}\left[Q+\frac{1+\gamma^2}{1-\gamma^2}+\frac{2\gamma}{1-\gamma^2}\star \mathcal{P}\right]\, ,
\end{equation}
and we obtain
\begin{align}\label{eq:constr1}
 &-\partial_\gamma\delta\hat{\mathcal{V}}\Big\lvert_{\gamma=-1}=\rho\,\frac{P_++\mathcal{P}_+}{\partial_+\rho+\mathcal{R}_+}= \rho\,\frac{P_--\mathcal{P}_-}{\partial_-\rho-\mathcal{R}_-}\, ,\\
 \label{eq:constr2}&-\partial_\gamma\delta\hat{\mathcal{V}}\Big\lvert_{\gamma=+1}=-\rho\,\frac{P_+-\mathcal{P}_+}{\partial_+\rho-\mathcal{R}_+}= -\rho\,\frac{P_-+\mathcal{P}_-}{\partial_-\rho+\mathcal{R}_-}\, .
\end{align}
Upon insertion of \eqref{eq:constr1} and \eqref{eq:constr2} into \eqref{eq:Residuesintegral}, we reach
\begin{equation}\label{eq:Virr1}
    \tilde{\mathcal{S}}_\pm=\pm\frac{\rho}{4}\text{Tr}\left(\frac{(P_\pm\pm\mathcal{P}_\pm)(P_\pm\pm\mathcal{P}_\pm)}{\partial_\pm\rho\pm\mathcal{R}_\pm}-\frac{(P_\pm\mp\mathcal{P}_\pm)(P_\pm\mp\mathcal{P}_\pm)}{\partial_\pm\rho\mp\mathcal{R}_\pm}\right)\, ,
\end{equation}
which is by no means the unique way to write it, given the non trivial relations \eqref{eq:constr1} and \eqref{eq:constr2}. Expressing everything in terms of auxiliary fields (which requires the imposition of the algebraic equations of motion \eqref{eq:algebraicP}, \eqref{eq:algebraicrho}), we attain a unique expression,
\be \label{eq:Virr2}
\tilde{\mathcal{S}}_\pm\doteq-\frac{1}{2}\rho\,\text{Tr}\left(\frac{v_\pm v_\pm}{\chi_{2\,\pm}}+ \frac{K_\pm{}^\mp v_\mp v_\mp}{\chi_{2\,\mp}}\right)\, .
\ee
\subsection{Recovery of equation \eqref{eq:Boxsigma+smth}}\label{appendix:BoxsigmafromVirasoro}
In this section, we prove that equation \eqref{eq:Boxsigma+smth}, written in terms of auxiliary fields
\be\label{eq:Boxsigma+smthaux}
\partial_+\mathcal{S}_-+\partial_-\mathcal{S}_+\doteq(-1+K_+{}^-K_-{}^+)\text{Tr}(v_+v_-)\, ,
\ee
stems from \eqref{eq:Virr2} with the aid of all the other field equations, and therefore is not an independent one. Let us begin from
\begin{align}
  \nonumber D_-\tilde{\mathcal{S}}_+=  \partial_-\tilde{\mathcal{S}}_+\doteq&-\left[\frac{1}{2}\partial_-\rho \text{Tr}\left(\frac{v_+v_+}{\chi_{2\,+}}+K_+{}^- \frac{v_-v_-}{\chi_{2\,-}}\right)+\rho\text{Tr}\left(\frac{v_+D_-v_+}{\chi_{2\,+}}-\frac{v_+v_+}{2\chi_{2\,+}^2}\partial_-\chi_{2\,+}\right)\right.\\[2pt]
  &\left.+\frac{\rho}{2}\text{Tr}\left(\frac{D_-(K_+{}^-v_-)v_-}{\chi_{2\,-}}\right)+\frac{\rho}{2}\text{Tr}\left(\frac{K_+{}^-v_-D_-v_-}{\chi_{2\,-}}\right)-\frac{\rho}{2}\text{Tr}\left(\frac{K_+{}^-v_-v_-\partial_-\chi_{2\,-}}{\chi_{2\,-}^2}\right)\right]\,\, .
\end{align}
We can manipulate the first of the two terms in the first parenthesis by means of equation \eqref{eq:dynamicaleom}, rewritten in terms of auxiliary fields,
\be
\partial_-\rho (v_++K_+{}^-v_-)+\partial_+\rho (v_-+K_-{}^+v_+)+\rho D_+ (v_-+K_-{}^+v_+)+\rho D_-(v_++K_+{}^-v_-)\doteq0\, ,
\ee
getting
\begin{align}
    \nonumber\partial_-\tilde{\mathcal{S}}_+\doteq&-\left[\frac{1}{2}(+1-K_+{}^-K_-{}^+)\text{Tr}(v_+v_-)-\frac{\partial_+\rho}{2\chi_{2\,+}}K_-{}^+\text{Tr}(v_+v_+)-\color{magenta}\frac{\rho}{2}\text{Tr}\left(\frac{D_-(K_+{}^-v_-)v_+}{\chi_{2\,+}}\right)\color{black}\right.\\[2pt]\nonumber -&\frac{\rho}{2}\text{Tr}\left(\frac{D_+(K_-{}^+v_+)v_+}{\chi_{2\,+}}\right)
     +\frac{1}{2}\frac{\partial_-\rho}{\chi_{2\,-}} K_+{}^-\text{Tr}(v_-v_-)+\color{magenta}\frac{\rho}{2}\text{Tr}\left(\frac{D_-v_+v_+}{\chi_{2\,+}}\right)\color{black}-\color{magenta}\frac{\rho}{2}\text{Tr}\left(\frac{D_+v_-v_+}{\chi_{2\,+}}\right)\color{black}\\[2pt]
    \nonumber \hspace{1cm} +&\left.\frac{\rho}{2}\text{Tr}\left(\frac{D_-(K_+{}^- v_-)v_-}{\chi_{2\,-}}\right)
    +\frac{\rho}{2}\text{Tr}\left(K_+{}^-\frac{D_-(v_-)v_-}{\chi_{2\,-}}\right)\right.\\[2pt]
     &\hspace{2.5cm} \left. -\frac{\rho}{2}\text{Tr}(v_+v_+)\frac{\partial_-\chi_{2\,+}}{\chi_{2\,+}^2}-\frac{\rho}{2}K_+{}^-\text{Tr}(v_-v_-)\frac{\partial_-\chi_{2\,-}}{\chi_{2\,-}^2} \right]\, .
\end{align}
The three terms in magenta recombine, via the Bianchi identities
\be
D_-v_+-D_+v_--D_-(K_+{}^-v_-)+D_+(K_-{}^+v_+)\doteq0\,, 
\ee
into $-\frac{\rho}{2}\text{Tr}\left(\frac{D_+(K_-{}^+v_+)v_+}{\chi_{2\,+}}\right)$, simplifying the expression to
\begin{align}
  \nonumber\partial_-\tilde{\mathcal{S}}_+\doteq&-\left[\frac{1}{2}(1-K_-{}^+K_+{}^-)\text{Tr}(v_+v_-)-\frac{\partial_+\rho}{2\chi_{2\,+}}K_-{}^+\text{Tr}(v_+v_+)+\frac{1}{2}\frac{\partial_-\rho}{\chi_{2\,-}} K_+{}^-\text{Tr}(v_-v_-)\right.\\[2pt]
    \nonumber &-\rho\text{Tr}\left(\frac{D_+(K_-{}^+v_+)v_+}{\chi_{2\,+}}\right)+\frac{\rho}{2}\text{Tr}\left(\frac{D_-(K_+{}^- v_-)v_-}{\chi_{2\,-}}\right)+\frac{\rho}{2}\text{Tr}\left(K_+{}^-\frac{D_-(v_-)v_-}{\chi_{2\,-}}\right)\\[2pt]
    &\left.-\frac{\rho}{2}\text{Tr}(v_+v_+)\frac{\partial_-\chi_{2\,+}}{\chi_{2\,+}^2}-\frac{\rho}{2}K_+{}^-\text{Tr}(v_-v_-)\frac{\partial_-\chi_{2\,-}}{\chi_{2\,-}^2}\right]\, .  
\end{align}
Completely analogous steps for  $\partial_+\tilde{\mathcal{S}}_-$ lead to
\begin{align}
 \nonumber\partial_+\tilde{\mathcal{S}}_-\doteq&- \left[\frac{1}{2}(1-K_-{}^+K_+{}^-)\text{Tr}(v_+v_-)+\frac{\partial_+\rho}{2\chi_{2\,+}}K_-{}^+\text{Tr}(v_+v_+)-\frac{1}{2}\frac{\partial_-\rho}{\chi_{2\,-}} K_+{}^-\text{Tr}(v_-v_-)\right.\\[2pt]
    \nonumber &-\rho\text{Tr}\left(\frac{D_-(K_+{}^-v_-)v_-}{\chi_{2\,-}}\right)+\frac{\rho}{2}\text{Tr}\left(\frac{D_+(K_-{}^+ v_+)v_+}{\chi_{2\,-}}\right)+\frac{\rho}{2}\text{Tr}\left(K_-{}^+\frac{D_+(v_+)v_+}{\chi_{2\,+}}\right)\\[2pt]
    &\left.-\frac{\rho}{2}\text{Tr}(v_-v_-)\frac{\partial_+\chi_{2\,-}}{\chi_{2\,-}^2}-\frac{\rho}{2}K_-{}^+\text{Tr}(v_+v_+)\frac{\partial_+\chi_{2\,+}}{\chi_{2\,+}^2}\right]\, .    
\end{align}
Summing the two terms, we obtain
\begin{align}
 \nonumber &\partial_+\tilde{\mathcal{S}}_-+\partial_-\tilde{\mathcal{S}}_+\doteq(-1+K_-{}^+K_+{}^-)\text{Tr}(v_+v_-)+\frac{\rho}{2}\text{Tr}\left(\frac{D_+(K_-{}^+v_+)v_+}{\chi_{2\,+}}\right)\\[2pt]
 \nonumber+&\frac{\rho}{2}\text{Tr}\left(\frac{D_-(K_+{}^-v_-)v_-}{\chi_{2\,-}}\right)-\frac{\rho}{2}\text{Tr}\left(K_-{}^+\frac{D_+(v_+)v_+}{\chi_{2\,+}}\right)-\frac{\rho}{2}\text{Tr}\left(K_+{}^-\frac{D_-(v_-)v_-}{\chi_{2\,-}}\right)\\[2pt]
 \nonumber  +&\frac{\rho}{2\chi_{2\,+}^2}\text{Tr}(v_+v_+)\left[\partial_-\chi_{2\,+}+K_-{}^+\partial_+\chi_{2\,+}\right]+\frac{\rho}{2\chi_{2\,-}^2}\text{Tr}(v_-v_-)\left[\partial_+\chi_{2\,-}+K_+{}^-\partial_-\chi_{2\,-}\right]\, ,
\end{align}
namely
\begin{align} \label{eq:preliminaryconstr}
   \nonumber &\partial_+\tilde{\mathcal{S}}_-+\partial_-\tilde{\mathcal{S}}_+\doteq(-1+K_-{}^+K_+{}^-)\text{Tr}(v_+v_-) \\[2pt]
   &+\frac{\rho}{2\chi_{2\,-}^2}\text{Tr}(v_-v_-)\left[\partial_+\chi_{2\,-}+\partial_-(K_+{}^-\chi_{2\,-})\right]+\frac{\rho}{2\chi_{2\,+}^2}\text{Tr}(v_+v_+)\left[\partial_-\chi_{2\,+}+\partial_+(K_-{}^+\chi_{2\,+})\right]\, ,
\end{align}
which can be further simplified. Indeed, equation \eqref{eq:Boxrho=smth} rewritten in terms of auxiliary fields reads,
\be
\partial_+\chi_{2\,-}+\partial_-\chi_{2\,+}+\partial_-(K_+{}^-\chi_{2\,-})+\partial_+(K_-{}^+\chi_{2\,+})\doteq0\, ,
\ee
which, combined with the condition $\epsilon^{\mu\nu}\partial_\mu\partial_\nu\rho=0$, which we here rewrite as
\begin{equation}
    \partial_+\chi_{2\,-}-\partial_-\chi_{2\,+}+\partial_-(K_+{}^-\chi_{2\,-})-\partial_+(K_-{}^+\chi_{2\,+})\doteq0\, ,
\end{equation}
leads to the additional relation
\be
\partial_+\chi_{2\,-}+\partial_-(K_+{}^-\chi_{2\,-})=\partial_-\chi_{2\,+}+\partial_+(K_-{}^+\chi_{2\,+})\doteq 0\, .
\ee
Plugged back into equation \eqref{eq:preliminaryconstr}, this condition returns the equation of motion \eqref{eq:Boxsigma+smthaux}
\begin{equation}
   \partial_+\tilde{\mathcal{S}}_-+\partial_-\tilde{\mathcal{S}}_+=(-1+K_-{}^+K_+{}^-)\text{Tr}(v_+v_-)\, ,
\end{equation}
for $\partial_+\tilde{\mathcal{S}}_-+\partial_-\tilde{\mathcal{S}}_+\doteq\partial_+\mathcal{S}_-+\partial_-\mathcal{S}_+$. Hence, $\tilde{\mathcal{S}}_\mu$ must be a specifically modified version of $\mathcal{S}_\mu$, (as it happens in the undeformed case for derivatives of $\sigma$) such that $\partial_\mu\tilde{\mathcal{S}}^\mu=\partial_\mu \mathcal{S}^\mu$.

\section{Yang-Baxter deformation of Principal Chiral Model coupled to $D=2$ dilaton-gravity}
\label{appendix:PCM2dgravity}

In order to connect the approach and notation of \cite{Hoare:2020fye} to the one in this present paper, we also present the Yang-Baxter deformation of the Principal Chiral Model as matter sector of a $D=2$ dilaton-gravity theory. The first subsection will be a review, as in \cite{Hoare:2020fye} the same was presented for the 'left' deformation. In the second subsection we present how to couple this to $D=2$ dilaton-gravity action.

\subsection{Review: Yang-Baxter deformation of Principal Chiral Model with spacetime dependent couplings}
The Lagrangian the standard (left) Yang-Baxter deformation \cite{Klimcik:2002zj} of the Principal Chiral Model is:
\begin{equation}
    L = \frac{1}{2} \rho(\tau,\sigma) \text{tr} \left( j_+ \frac{1}{1 - \eta(\tau,\sigma) R_g} j_- \right)
\end{equation}
where $j = g^{-1} \mathrm{d} g$ and $\rho$ and $\eta$ would be the coupling constant and deformation parameter respectively for the standard $\sigma$-model picture. They are promoted to fields in $D=2$.\footnote{At this stage $\rho$ and $\eta$ are not considered to be physical fields.} Moreover, as in the main text, we have that $R_g = \text{Ad}_{g}^{-1} \circ R \circ \text{Ad}_g$, where $R: \ \mathfrak{g} \rightarrow \mathfrak{g}$ is a solution to the (modified) classical Yang-Baxter equation:
\begin{equation}
    [R(m),R(n)] - R([R(m),n] - [m,R(n)]) = - c^2 [m,n]
\end{equation}
where $c^2 = 0,1,-1$ (other values can be reabsorbed into a redefinition of the deformation parameter $\eta$).

The equations of motion become
\begin{equation}
    \mathrm{d} \star (\rho k) = 0 \quad \rightarrow \quad \mathrm{d} \star k = - \frac{\mathrm{d} \rho}{\rho} \wedge \star k
\end{equation}
with
\begin{equation}
    j = k + \eta \star R_g(k).
\end{equation}
In terms of these currents, the non-trivial character of $\eta$ only shows in the rewriting of the Maurer-Cartan equation of $j$ in terms of $k$.
\begin{align}
    \mathrm{d} j + \frac{1}{2} [j,j] &= \mathrm{d} k + \frac{1}{2} (1 - c^2 \eta^2) [k,k] + \left(\mathrm{d} \eta - \eta \frac{\mathrm{d}\rho}{\rho} \right) \wedge \star R_g (k) = 0 \label{eq:YBPCMMaurerCartan}.
\end{align}
A simplifying assumption is
\begin{align}
    \mathrm{d} \eta - \eta \frac{\mathrm{d}\rho}{\rho} = 0\, .
\end{align}
This means that $\eta = 0$ is a fixed point that corresponds to the known undeformed PCM (or coset) coupled to the $\rho$. In the following, we assume that $\eta \neq 0$ which in turns implies that the dynamics of $\eta$ and $\rho$ are essentially equivalent
\begin{align}
     \frac{\mathrm{d}\eta}{\eta} = \frac{\mathrm{d}\rho}{\rho}.
\end{align}
This means that even for the homogeneous Yang-Baxter deformation ($c=0$), where the classical dynamics is equivalent to the undeformed case, the deformation parameter can receive non-trivial dynamics, analogously to $\rho$.

Making a similar ansatz for the Lax pair as in the case with constant couplings, we consider:
\begin{align}
    \mathfrak{L}(\lambda) = (1 - c^2 \eta^2) \left( \frac{1}{1 - \lambda^2} k  + \frac{\lambda}{1 - \lambda^2} \star k \right).
\end{align}
The Lax connection is flat, given that $k$ satisfies equation of motion, Maurer-Cartan identity and the spectral parameter is dynamical and subject to the following condition:
\begin{equation}
    \frac{\mathrm{d} \lambda}{\lambda} - \frac{2 \lambda}{1 + \lambda^2} \frac{\star \mathrm{d} \lambda}{\lambda} = \frac{1 + c^2 \eta^2}{1 - c^2 \eta^2} \frac{1 - \lambda^2}{1 + \lambda^2} \frac{\mathrm{d} \eta}{\eta} - \frac{2 c^2 \eta^2}{1 - c^2 \eta^2} \frac{1 - \lambda^2}{(1 + \lambda^2 ) \lambda} \frac{\star \mathrm{d} \eta}{ \eta}
\end{equation}
or equivalently:
\begin{equation}
    \frac{\mathrm{d} \lambda}{\lambda} = \frac{1}{(1 - c^2 \eta^2)(1 - \lambda^2)} \left( \left( (1 + \lambda^2)(1 + c^2 \eta^2) - 4 c^2 \eta^2 \right) \frac{\mathrm{d}\eta}{\eta} + \frac{2 ( \lambda^2 - c^2 \eta^2 )}{\lambda} \frac{\star \mathrm{d}\eta}{\eta} \right).
\end{equation}
Integrability of this requires
\begin{equation}
    \mathrm{d} \star \mathrm{d} \eta + \frac{4c^2 \eta}{1 - c^2 \eta^2} \mathrm{d} \eta \wedge \star \mathrm{d}\eta = 0.
\end{equation}
This reproduces the form (3.10) of \cite{Hoare:2020fye}
\begin{equation}
    \mathrm{d} \star \mathrm{d} \eta - \frac{\beta^\prime(\eta)}{\beta(\eta)} \mathrm{d} \eta \wedge \star \mathrm{d} \eta = 0,
\end{equation}
for $\beta = (1 - c^2 \eta^2)^2$, which is in fact the $\beta$-function for the coupling $\eta$ \cite{Hoare:2019ark}.
This implies that in suitable coordinates $\eta$ can have non-trivial dynamics along the RG-flow with world-sheet time being RG-time. The dynamics of $\rho$ is then related through $\mathrm{d} \ln \eta = \mathrm{d} \ln \rho$. This implies $\eta = C \rho$, with $C$ now playing the role of the deformation parameter.

\subsection{Coupling to $D=2$ dilaton-gravity}
Let us present in the following, how other integrable models considered in \cite{Hoare:2020fye} could be coupled to $D=2$ dilaton gravity. Also, we motivate the form of the action \eqref{eq:YB2dGrav}, with the function $\beta(\rho)$ in the $D=2$ gravity part of the action.

Adding the coupling to 2d dilaton-gravity to this $\sigma$-model with space-time dependent couplings leads to:
\begin{equation}
    L = (\partial_+ \rho \partial_- \sigma + \partial_- \rho \partial_+ \sigma) - \frac{1}{2} \bar{\rho} \text{tr} \left( j_+ \frac{1}{1 - \eta R_g} j_- \right)
\end{equation}
in addition to potential Virasoro constraints. Following the calculation above, integrability of the matter sector (the $\sigma$-model part) requires:
\begin{align*}
    0 &= \mathrm{d} \star \mathrm{d} \eta - \frac{\beta^\prime(\eta)}{\beta(\eta)} \mathrm{d} \eta \wedge \star \mathrm{d} \eta, \\
    0 &= \mathrm{d} \left( \frac{\eta}{\bar{\rho}} \right), \qquad \Rightarrow \qquad \bar{\rho} = C \eta\, .
\end{align*}
The dynamics of $\sigma$ (i.e. the equation of motion from varying w.r.t. to $\rho$), is not reproduced from the Lax connection with spacetime dependent spectral parameter and we neglect it here. This equation of motion has the same form as in the coset case \eqref{eq:YBEquationOfMotionSigma}. As is the undeformed case, one would expect that the equations of motion for $\sigma$ would follow from a combination of the Virasoro constraints and the equations of motion for $\rho$ and the matter part. As we consider this model as coming from a higher dimensional gravity theory, the derivation of $D=2$ Virasoro constraints should be related to this higher-dimensional origin. This higher-dimensional origin is not at all clear for the Yang-Baxter deformation, as discussed in the main text. Hence, the derivation of the Virasoro constraints goes beyond the scope of the present paper, and we do not consider the dynamics of $\sigma$ here.

These conditions imply
\begin{equation}
    \mathrm{d} \star \mathrm{d} \bar{\rho} - \frac{\bar{\beta}^\prime(\bar{\rho})}{\bar{\beta}(\bar{\rho})} \mathrm{d} \bar{\rho} \wedge \star \mathrm{d} \bar{\rho} = 0
\end{equation}
for $\bar{\beta}(\bar{\rho}) = \beta\left( \frac{\bar{\rho}}{C}\right)$. In order to relate to $\mathrm{d}\star \mathrm{d} \rho = 0$ (from the first term in the Lagrangian), $\rho$ and $\bar{\rho}$ have to be related as follows:
\begin{equation}
    \mathrm{d} \rho = \frac{\mathrm{d} \bar{\rho}}{\bar{\beta}(\bar{\rho})}.
\end{equation}
In this case $\beta = (1 - c^2 \bar{\rho}^2)^2$. This can be integrated to:
\begin{equation}
    \rho = \frac{1}{2} \left( \frac{\bar{\rho}}{1 - \frac{c^2}{C^2} \bar{\rho}^2} + \frac{\tanh^{-1} \left( \frac{c}{C} \bar{\rho}\right)}{\frac{c}{C}} \right) + C_2
\end{equation}
for some integration constants $C$ and $C_2$.
\newpage
\bibliographystyle{JHEP}
\bibliography{Bib}

\end{document}